\documentclass[12pt]{article}

\pdfoutput=1

\usepackage{color}
\usepackage{amsmath}
\usepackage{amsfonts}
\usepackage{amssymb}
\usepackage{caption}
\usepackage{graphicx}
\usepackage{slashed}            % for slashed characters in math mode
\usepackage{subcaption}			% subcaption works better with hyperref
\usepackage{xspace}				% For spacing after commands
\usepackage{cite}
\usepackage{booktabs}

\usepackage[english]{babel}
\usepackage{fancyhdr}
\usepackage{amsmath}
\usepackage{amssymb}
\usepackage{amsfonts}
\usepackage{psfrag}
\usepackage[applemac]{inputenc}
\usepackage{graphicx}
\usepackage{bm}
\usepackage{multirow}
\usepackage[	colorlinks=true,	citecolor=black,	linkcolor=black,	urlcolor=blue,hypertexnames=false]{hyperref}

\def\gsim{\mathrel{\rlap{\lower4pt\hbox{\hskip1pt$\sim$}}
    \raise1pt\hbox{$>$}}}                % greater than or approx. symbol
\newcommand{\arXiv}[2]{\href{http://arxiv.org/pdf/#1}{{\tt #2/#1}}}
\newcommand{\arXivold}[1]{\href{http://arxiv.org/pdf/#1}{{\tt #1}}}

\newcommand{\beq}{\begin{eqnarray}}
\newcommand{\eeq}{\end{eqnarray}}

\newcommand{\bag}{\begin{align}}
\newcommand{\eag}{\end{align}}

\newcommand{\MeV}{\,\mathrm{MeV}}

\begin{document}
\begin{titlepage}

\vskip.5cm

\begin{center} %TITLE HERE
{\huge \bf Neutron Star Mergers Chirp About}
 \vskip.15cm
{\huge \bf Vacuum Energy} 
\end{center}
\vspace*{0.4cm} 

\begin{center} % AUTHORS HERE
{\bf \   Csaba Cs\'aki$^a$, Cem Er\"{o}ncel$^b$, Jay Hubisz$^b$, \\Gabriele Rigo$^b$, and John Terning$^c$} 
\end{center}
%\vskip 4pt

\begin{center} 
% PLACES HERE
 
$^{a}$ {\it Department of Physics, LEPP, Cornell University, Ithaca, NY 14853, USA} \\

\vspace*{0.1cm}

$^{b}$ {\it Department of Physics, Syracuse University, Syracuse, NY  13244, USA} \\

\vspace*{0.1cm}

$^{c}$ {\it Department of Physics, University of California, Davis, CA 95616, USA} \\

\vspace*{0.1cm}

{\tt  
 
\href{mailto:csaki@cornell.edu}{csaki@cornell.edu}, 
 \href{mailto:ceroncel@syr.edu}{ceroncel@syr.edu}, \href{mailto:jhubisz@syr.edu}{jhubisz@syr.edu},  
  \\ \href{mailto:garigo@syr.edu}{garigo@syr.edu},
 \href{mailto:jterning@gmail.com}{jterning@gmail.com}}

\end{center}

\vglue 0.3truecm

\centerline{\large\bf Abstract}
\begin{quote}
Observations of gravitational waves from neutron star mergers open up novel directions for exploring fundamental physics: they offer the first access to the structure of objects with a non-negligible contribution from vacuum energy to their total mass. The presence of such vacuum energy in the inner cores of neutron stars occurs in new QCD phases at large densities, with the vacuum energy appearing in the equation of state for a new phase. This in turn leads to a change in the internal structure of neutron stars and  influences their tidal deformabilities which are measurable in the chirp signals of merging neutron stars.  By considering three commonly used neutron star models we show that for large chirp masses the effect of vacuum energy on the tidal deformabilities can be sizable.  Measurements of this sort have the potential to provide a first test of the gravitational properties of vacuum energy independent from the acceleration of the Universe, and to determine the size of QCD contributions to the vacuum energy. 
 \end{quote}

\end{titlepage}

%\newpage

\setcounter{equation}{0}
\setcounter{footnote}{0}

%%%%%%%%%%%%%%%%%%%%%%%%%%%%%%%%%%%%%%%%%%%%%%%%%%%%%%%%%%%%%%%%%%%%%%%%%%%%%%%%%%%
%%%%%%%%%%%%%%%%%%%%%%%%%%%%%%%%%%%%%%%%%%%%%%%%%%%%%%%%%%%%%%%%%%%%%%%%%%%%%%%%%%%
\section{Introduction}
\setcounter{equation}{0}

The recent observations of gravitational waves (GW's) from the merger of neutron stars (NS's) by LIGO/Virgo~\cite{TheLIGOScientific:2017qsa} along with the corresponding electromagnetic observations of the resulting kilonova have reverberated across most areas of physics and astronomy. From the point of view of particle physics the most important consequence of GW170817 and future merger events is our new ability to directly examine the properties of the QCD matter forming the inner layers of NS's, allowing us to use NS's as laboratories for fundamental physics \cite{Baryakhtar:2017dbj,Croon:2017zcu,Ellis:2017jgp}. This might also open up new avenues to testing the gravitational properties of vacuum energy (VE) which may also get at the heart of some of the deepest puzzles in fundamental physics~\cite{Bellazzini:2015wva}. 

It has long been speculated that there may be a new phase of nuclear matter at the core of the NS's~\cite{Ivanenko}. If such a phase indeed exists it is expected to be accompanied by a jump in VE \cite{Ellis:1991qx} of order $\Lambda_\textnormal{QCD}^4$ (where $\Lambda_\textnormal{QCD} \sim 200\MeV$ is the usual QCD scale)  making NS's the only known objects where VE might make up a non-negligible fraction of the total mass. Therefore studies of the interior structure of NS's can also probe the gravitational properties of VE, possibly shedding light on some of the most interesting open questions in physics: it could provide verification of the equivalence principle for VE. This would be the first test independent of those obtained from the cosmic acceleration of the Universe. Acceleration of the Universe provides information on VE in the low-temperature low density phase of the SM, while NS's could probe a low temperature but very high density phase if it exists in their cores. This could allow us to isolate the QCD contribution to ordinary VE and probe its gravitational properties.

Alongside the exciting advent of the gravitational wave observation era shepherded in by  LIGO/Virgo, the Neutron star Interior Composition ExploreR  (NICER) mission will soon measure masses and radii of several millisecond pulsars~\cite{Bogdanov1610404B}. These measurements as well as the chirps from the inspiral of merging neutron stars can provide information about the equation of state (EoS) of dense nuclear matter. The chirps in particular are sensitive to the tidal deformability  of NS's as they approach each other~\cite{Flanagan:2007ix,Hinderer:2007mb,Hinderer:2009ca,Postnikov:2010yn,Lackey:2014fwa}. There has already been considerable work on constraining the EoS  using the new LIGO/Virgo data~\cite{Margalit:2017dij,Bauswein:2017vtn,Zhou:2017xhf,Rezzolla:2017aly,Annala:2017llu}.

There exists an extensive literature focused on trying to put bounds on the nuclear EoS at high densities from neutron star measurements (see for example \cite{Lattimer:2000nx,Douchin:2001sv,Read:2008iy,Hebeler:2013nza}).
Some recent theoretical work has focused on modeling possible new phases at the cores of neutron stars  by using quasi-particle quarks  rather than neutrons to provide the simplest description of the microscopic physics~\cite{Mueller:1996pm,Alford:2007xm,Agrawal2010,Bejger:2016emu,Alvarez-Castillo:2017qki,Alford:2017qgh,Alford:2013aca}.
Further work has been done using NS's to constrain ``beyond the Standard Model" physics~\cite{Baryakhtar:2017dbj,Croon:2017zcu,Ellis:2017jgp}. 

In this paper we will assume that there is only Standard Model physics involved in the composition of neutron stars, and we will not try to model the microphysics of the putative new phase. Our main goal is to investigate the observable effects of the presence of VE on the GW signal as well as the mass versus radius curve of NS's, possibly providing new experimental probes of VE. To achieve this we will  parameterize the effect of the new phase with a jump in the ground state energy due to a QCD phase transition assumed at the core~\cite{Bellazzini:2015wva,Alford:2013aca,Seidov71,Shaeffer83}. This new phase would appear at a critical pressure of order $p_\textnormal{c} \propto \Lambda^4_\textnormal{QCD}$, and is expected to also lead to a change of VE \cite{Ellis:1991qx} of order $\Delta \Lambda \propto \Lambda_\textnormal{QCD}^4$. We will follow the conventional models for NS's where the EoS is divided into 7 layers, but we modify the innermost layer to take the effect of VE at the core into account. We will then evaluate the tidal Love numbers for such models. We stress that the existence of the new phase requires a modification of the traditional assumptions used in NS simulations regarding  energy density jumps at the boundary of a new phase  \cite{Alford:2013aca,Seidov71,Shaeffer83}.  We will explore the effect of a difference in energy densities of the two phases that includes a discontinuous, density independent term reflecting the absence of the low density QCD contribution to VE \cite{Bellazzini:2015wva}.  We will present several models of NS cores and estimate the effect of VE on tidal Love numbers. We find that VE can have a significant effect on NS merger waveforms with high chirp masses, so that such events serve as a probe of the physics of vacuum energy.

The paper is organized as follows. In Section 2 we present the models we use for nuclear matter in the interior of NS's, along with a detailed discussion of the treatment of the phase transition at the boundary of the innermost layer. Section 3 contains the description of the tidal deformability of NS's. The results of our simulations and the effects of VE on the NS observables are given in Section 4: we show the mass versus radius curves and the tidal deformabilities for three different well-studied NS models and the effects of VE on those observables. Finally we conclude in Section 5. 

%%%%%%%%%%%%%%%%%%%%%%%%%%%%%%%%%%%%%%%%%%%%%%%%%%%%%%%%%%%%%%%%%%%%%%%%%%%%%%%%%%%
%%%%%%%%%%%%%%%%%%%%%%%%%%%%%%%%%%%%%%%%%%%%%%%%%%%%%%%%%%%%%%%%%%%%%%%%%%%%%%%%%%%
\section{Modeling High Density QCD \label{sec:ns}}
\setcounter{equation}{0}
\setcounter{footnote}{0}
The main difference between our work and that of previous studies of tidal deformability of NS's is that we will fully account for a phase transition to an exotic phase of QCD in the innermost core region of NS's.  Crucially, we take into account the Standard Model expectation that there is a constant shift $\Lambda$, independent of baryon number density, in the ground state energy relative to the surrounding layers parametrizing the change in VE due to the phase transition. In the ordinary phase of QCD the nonperturbative condensates of quarks and gluons make contributions \cite{Ellis:1991qx} of order $(100\,{\rm MeV})^4$ to the VE.  These contributions, along with those from other sectors of the SM, are canceled 
by the ``bare" cosmological constant 
  down to the observed cosmological constant of order $({\rm meV})^4$:
\begin{equation}
\Lambda_{{\rm QCD}}^{{\rm SM \, vac}}+\Lambda_{{\rm SM\, other}}+ \Lambda_{{\rm bare}} \simeq (10^{-3}\ {\rm eV})^4\ .
\end{equation}
The origin of the mechanism leading to this cancelation remains unknown.   In an exotic phase of QCD the QCD contributions to the VE will have order one modifications and hence the precise cancelation will no longer apply:
\begin{equation}
\Lambda_{{\rm QCD}}^{{\rm exotic}}+\Lambda_{{\rm SM\, other}}+\Lambda_{{\rm bare}} \simeq \Delta \Lambda \ ,
\end{equation}
where $\Delta \Lambda $ is the shift in the QCD vacuum energy due to the phase transition. 
Hence in the absence of a dynamical adjustment mechanism, Standard Model physics predicts a density independent shift in the energy of the exotic phase compared to the ordinary phase, which will serve as a new effective cosmological constant term for this phase.   An estimate of the difference between the VE of the exotic phase and the ordinary vacuum is given by nuclear saturation density: $|\Delta \Lambda | \sim \Lambda_\textnormal{QCD}^4$ \cite{Bellazzini:2015wva}.  Such a phase change is strongly suspected to occur at high chemical potential, with theoretical evidence arising from truncated diagrammatic expansions and other approximate methods \cite{Alford:2007xm,Agrawal2010}.  The phase change is in fact  part of the standard picture of the QCD phase diagram.  For many plausible descriptions of the matter in the outer portions of the star, nuclear saturation density is approached near the core of the densest NS's, making it quite possible that the most massive NS's contain cores with an exotic phase.  In this section, we give a description of how one can model the QCD equation of state at various pressures, with particular attention paid to the phase transition that may occur in the innermost region.

%%%%%%%%%%%%%%%%%%%%%%%%%%%%%%%%%%%%%%%%%%%%%%%%%%%%%%%%%%%%%%%%%%%%%%%%%%%%%%%%%%%
\subsection{Modeling the Outer Layers}
The physics of neutron stars is an extremely rich field, and there are many details that go into modeling the different regions of NS's.  Such an analysis is well beyond the scope of this work, however there are methods for coarse-graining these complexities to obtain an approximate equation of state for nuclear matter up until the phase transition we are interested in.  Such an approximation is sufficient for the purposes of making predictions for gravitational wave signals.  The most common methodology for modeling  the high density nuclear physics region outside the exotic phase core is to separate the neutron star into multiple layers, with each layer satisfying a non-relativistic polytropic equation of state. The parameters of the polytrope are fixed either by matching conditions or by fitting results from more detailed studies.

We follow this established methodology and model the nuclear fluid and its corresponding EoS as a piecewise polytrope where the boundaries between each layer are set by a given value of the pressure. Following previous work \cite{Read:2008iy,Hebeler:2013nza} we will parametrize the EoS with a total of 7 layers. The Israel junction conditions \cite{Israel} require that the pressure must always be continuous between layers, even if each side of the boundary is separated by  a first order phase transition.
It is traditional to parameterize the EoS by assuming that  the pressure is given by a power of the mass density $\rho(r)= m_n n(r)$ rather than a power of the energy density (as would be natural for a high-density, relativistic fluid). Since we want to efficiently compare our results with the existing state of the art simulations (some of which have been used as benchmarks for the LIGO/Virgo analysis) we will bow to this tradition and parametrize the EoS as 
\beq
 p =  K_{i} \rho^{\gamma_{i}}~,\quad \quad  p_{i-1} \leq p  \leq p_{i} ~,
\label{eq:polytropic}
\eeq
where $i \in \lbrace1, \ldots , 7\rbrace$ for $K_i, \gamma_i$ and $i \in \lbrace1, \ldots , 6\rbrace$ for $p_i$. 
The pressures, $p_i$, dividing the various layers have a one to one correspondence with the boundaries in the mass density: $\rho_i$. 
The Einstein equations contain the energy density, which is related to the mass density via the first law of thermodynamics: $d(\epsilon/\rho)=-p\,d(1/\rho)$. Integrating the first law together with (\ref{eq:polytropic}) yields the corresponding energy density:
\beq
\epsilon = (1+a_i)\rho + \frac{K_i}{\gamma_i-1} \rho^{\gamma_i}~,
\eeq
where the $a_i$ are integration constants. Note that the appearance of the $a_i$ parameters is a consequence of using a polytropic ansatz for the mass density. Naively, one would think that using a relativistic polytropic ansatz for the energy density would have led us to a relation with one less free parameter.  However another thermodynamical condition, continuity of the chemical potential, would have forced us to reintroduce the baryon number density, and therefore to bring back another parameter. So these simply correspond to different parametrizations of the EoS, and we adopt the one described above in order to follow the traditional approach. 

By using 7 layers we have introduced a large number of parameters ($\gamma_i, K_i$ and $a_i$). Most of those can be determined by continuity of various quantities at the layer boundaries. For the outer 6 layers we assume the continuity of the energy density at the boundaries, which allows us to determine the $a_i$'s:
\beq
a_i= \frac{\epsilon(\rho_{i-1})}{\rho_{i-1}}-1 -  \frac{K_i}{\gamma_i-1}\rho_{i-1}^{\gamma_i-1}~.
\label{aeq}
\eeq
If the $K_1$ constant for the outermost layer is known, then the other $K_i$ values (except for the innermost layer) can be determined by the continuity of the pressure:
\beq
K_i=K_{i-1}\rho_{i-1}^{\gamma_{i-1}-\gamma_{i}}~,\quad \quad i \in \lbrace 2,\ldots, 6\rbrace~.
\label{fixingK}
\eeq
For the outermost layer, the ``crust", we have $p_0=0$. Requiring that $\lim_{\rho\to 0} \frac{\epsilon}{\rho} =1$ (physically this means that the edge of the star is ordinary non-relativistic matter) implies that $a_1=0$. Thus the parameterization of the EoS of the NS for the outer layers will require us to specify the critical pressures $p_i$, all the polytropic exponents $\gamma_i$ as well as the outermost polytropic constant $K_1$, while all other parameters will be determined by the continuity conditions.

%%%%%%%%%%%%%%%%%%%%%%%%%%%%%%%%%%%%%%%%%%%%%%%%%%%%%%%%%%%%%%%%%%%%%%%%%%%%%%%%%%%
\subsection{Modeling the Core and the Effect of VE}

For the last layer, we use an equation of state that incorporates physics associated with a change in the QCD vacuum state due to high density.  There are two effects expected at this phase transition:  a vacuum energy term $\Lambda$ in the fluid that is independent of baryon number density, and a jump in energy density across the boundary.  Unlike in the outer layers, in the exotic phase the nature of the baryonic states may be very different from the usual zero temperature baryons. Since QCD conserves baryon number,  for the innermost layer it is more natural to use baryon number density $n$ in place of the mass density as the variable parametrizing the EoS for the central core  ($p>p_6$). In this case, the equation of state can be written as:
\begin{align}
p &= \tilde{K}_{7} n^{\gamma_{7}} - \Lambda~,\\
\epsilon &= \tilde{a}_7 n  + \frac{\tilde{K}_7}{\gamma_7-1} n^{\gamma_7}+ \Lambda ~.
\label{innnerlaayer1}
\end{align}
Note that the vacuum energy appears with the opposite sign in the energy density and pressure, just as with the cosmological constant.
Our goal is to see how sensitive neutron star observables are to the VE shift $\Lambda$. 

To keep the form of the EoS unchanged in the various layers we can introduce the density $\rho = m_n n$ where $m_n$ is the ordinary neutron mass, and use this rescaled number density for the innermost layer. We can easily see that in terms of this rescaled density the EoS will have the same form as for the outer layers:  
\begin{align}
p &= K_{7} \rho^{\gamma_{7}} - \Lambda~,\\
\epsilon &= (1+a_7) \rho  + \frac{K_7}{\gamma_7-1} \rho^{\gamma_7}+ \Lambda ~,
\label{innnerlaayer}
\end{align}
where $K_7= \tilde{K}_7 / m_n^{\gamma_7}, (1+a_7) = \tilde{a}_7/ m_n$ are just redefinitions of the unknown constants parametrizing the EoS for the inner layer. We adopt this notation in order to stay close to the standard formalism used  in the literature. 
 
Let us now examine in detail the continuity (or jump) of the various quantities at the phase boundary between the sixth and the seventh (innermost) layer. The Israel junction conditions \cite{Israel} still require that the pressure be continuous:
\beq
K_{7} \rho_+^{\gamma_{7}} - \Lambda= K_{6}\rho_-^{\gamma_{6}}=p_6~,
\label{continuity-pressure}
\eeq
but due to the appearance of the $\Lambda$ term this now requires a jump in $\rho(r)$ from $\rho_+$ to $\rho_-$ (where $\rho_-=\rho_6$) and  consequently also in $\epsilon(r)$ from $\epsilon_+$ to $\epsilon_-$. Since QCD conserves baryon number, another quantity that we need to require to be continuous is the chemical potential $\mu$ (that is we are assuming chemical equilibrium at the phase boundaries with conserved baryon number). The chemical potential at zero temperature is given by
\beq
\mu=\frac{\epsilon + p}{n}~,
\eeq
where $n$ is again the baryon number density.  This relation holds even if the VE is nonzero. Therefore the jumps from $\epsilon_+$ to $\epsilon_-$ and from $\rho_+$ and $\rho_-$ (in our convention $\rho_+ = m_n n_+$ and $\rho_- = m_n n_-$) are related to each other by
\beq
\frac{\epsilon_+ +p_6}{\rho_+}=\frac{\epsilon_- + p_6}{\rho_-}~.
\label{continuity-chempot}
\eeq
The convexity of the free energy $\left( \frac{\partial^2 F}{\partial V^2} \right)_{T,N} >0$ can be translated to  $\left( \frac{\partial p}{\partial n} \right)_{T,N} >0$. This latter form implies that the number density increases with pressure, yielding $\rho_+\geq\rho_-$. This condition together with the continuity of the chemical potential tells us that the jump in energy density should also be positive, i.e.\ $\epsilon_+\geq\epsilon_-$.

A typical phase transition will have both $\Delta \epsilon$ and $\Lambda$ non-vanishing, and this scenario is the focus of our studies.  
We choose to parametrize the jump in energy density such that it is proportional to the absolute value of the shift in VE:
\beq
\epsilon_+-\epsilon_-=\alpha\lvert\Lambda\rvert~.
\label{jump}
\eeq
For each value of $\gamma_7$, $\alpha$, and $\Lambda$, this condition, together with continuity of the chemical potential, fixes the values of $K_7$ and $a_7$. This parametrization of the phase transition has the advantage that the $\Lambda=0$ limit reproduces the results obtained in the literature since both the mass density and the energy density become continuous in this case.  In principle, $\alpha$ could be taken to be zero, isolating the effects of vacuum energy from a jump in the energy density, corresponding to a second order phase transition. Here we will assume that the phase transition is first order with an accompanying jump in most quantities across the phase boundary and take $\alpha > 0$.
A final consistency condition is that both the full pressure and the partial pressure of the fluid, $K_{7} \rho^{\gamma_{7}}$, must be positive.  This implies that $\Lambda$ must satisfy $-p_6 < \Lambda$.

%%%%%%%%%%%%%%%%%%%%%%%%%%%%%%%%%%%%%%%%%%%%%%%%%%%%%%%%%%%%%%%%%%%%%%%%%%%%%%%%%%%
%%%%%%%%%%%%%%%%%%%%%%%%%%%%%%%%%%%%%%%%%%%%%%%%%%%%%%%%%%%%%%%%%%%%%%%%%%%%%%%%%%%

\section{Modeling Neutron Stars\label{sec:GW}}
\setcounter{equation}{0}
\setcounter{footnote}{0}

After presenting the relevant physics of the dense QCD matter forming the interior of the NS we are now ready to review the usual method for calculating the structure of the interior of the NS.  GW emission observed by LIGO/Virgo originates from the inspiral phase, when the stars are far apart relative to their radii.  In this stage of the merger, the NS's are still well approximated by nearly spherically symmetric static objects, with deviations described by a linear response in an expansion in spherical harmonics. In this paper we will ignore the effects of NS angular momentum but plan to further investigate that in a future publication. First we briefly review the equations relevant for the spherically symmetric solution and then present an overview of the perturbations due to the gravitational field of the other NS. 

%%%%%%%%%%%%%%%%%%%%%%%%%%%%%%%%%%%%%%%%%%%%%%%%%%%%%%%%%%%%%%%%%%%%%%%%%%%%%%%%%%%
\subsection{Spherically Symmetric Solutions}

At lowest order, the stars are spherically symmetric, and their mass distribution is given by the solution to the Tolman-Oppenheimer-Volkoff (TOV) equations~\cite{Oppenheimer:1939ne}.
These equations are easily derived by starting with a spherically symmetric metric ansatz
\beq
ds^2=e^{\nu(r)}dt^2- \left(1-\frac{2 G m(r)}{r}\right)^{-1}dr^2-r^2d\Omega^2~,
\label{sphericalmetric}
\eeq
and using the associated Einstein equations assuming a spherically symmetric fluid distribution with pressure $p(r)$ and energy density $\epsilon(r)$.  The resulting TOV equations are:
\begin{align}
m^\prime (r)&= 4\pi r^2 \epsilon(r)~, \label{meq} \\ 
p^\prime(r)&=  -\frac{p(r)+\epsilon(r)}{r\left(r-2G m(r) \right)}G\left[m(r)+4\pi r^3 p(r)\right]~, \label{peq} \\
\nu^\prime(r)&= -\frac{2 p^\prime(r)}{p(r)+\epsilon(r)}~, \label{nueq}
\end{align}
where $^\prime$ denotes differentiation with respect to the radial coordinate $r$. The TOV metric provides the unperturbed solution around which the gravitational field of the second star will introduce perturbations that can be dealt with using a multipole expansion.  From the solution to these equations, one obtains the internal structure of the star:  the mass as a function of radius, as well as the thicknesses and masses of the various layers.

%%%%%%%%%%%%%%%%%%%%%%%%%%%%%%%%%%%%%%%%%%%%%%%%%%%%%%%%%%%%%%%%%%%%%%%%%%%%%%%%%%%
\subsection{Tidal Distortion and Love Numbers\label{sec:tidal}}

In a neutron star binary, each neutron star experiences gravitational tidal forces due to the other.  This force squeezes the stars along the axis passing through both of their centers, and deforms the stars, inducing a quadrupole moment.  The size of this induced quadrupole moment is determined by the structure of each neutron star, which can be  characterized by its compactness and the stiffness of the EoS. These in turn depend on the physical properties of the dense QCD matter as described  by its EoS discussed in the previous section.  The effect of the induced quadrupole on gravitational wave data is to change the power emission as a function of time and frequency.   Thus LIGO data on NS inspirals contains information about this tidal deformability, which depends on the equation of state of the matter making up the stars.

A common way to describe the deformability of a star is through the Love number.  Love numbers were originally introduced in the study of Newtonian tides \cite{Love}.  The application of Love numbers to gravitational waves produced in neutron star inspirals was initiated in refs. \cite{Flanagan:2007ix,Hinderer:2007mb}, and further generalized in \cite{Damour:2009vw}. Detailed studies of the prospects for gravitational wave detection were provided in \cite{Hinderer:2009ca,Postnikov:2010yn,Lackey:2014fwa}.

In the local rest frame of one star a small tidal field can be described in terms of a Taylor expansion of the Newtonian gravitational potential, or the time-time component of the metric tensor.  There are two contributions, one from the effect of the distant star, and the other from the induced quadrupole moment.
At large distances (using Cartesian coordinates, $x^i$)   $g_{tt}$ takes the form \cite{Hinderer:2009ca}
\beq
\frac{1+ g_{tt}}{2} \approx \frac{GM}{r}+\frac{3G Q_{ij}}{2 r^5}  x^i x^j - \frac{1}{2} {\mathcal E}_{ij} x^i x^j \ldots
\label{asympmetric}
\eeq
Here ${\mathcal E}_{ij}$ parametrizes the external tidal gravitational field, and $Q_{ij}$ is the induced quadrupole moment.  Both matrices are traceless and symmetric.  To linear order in the response,  the induced quadrupole is determined by the tidal deformability, $\lambda$, defined by
\beq
Q_{ij}=-\lambda \,{\mathcal E}_{ij}~.
\eeq
One can then define a dimensionless quantity $k_2$ by
\beq
k_2 =\frac{3}{2} \frac{G\lambda }{R^5}~,
\eeq
where $R$ is the radius of the neutron star. This is referred to as the $l=2$ tidal Love number, and is the main physical observable.  The advantage of this parametrization is that the Love number does not vary much with the size of the star, with typical Love numbers ranging from $k_2 = 0.001$ to $k_2=1$ as masses and equations of state are varied.

In order to determine $k_2$, one performs the perturbative expansion of the solutions to the Einstein equations in the presence of an external gravitational field assuming a multipole expansion.
Thus inside and near the star we will write the metric perturbation as an expansion in spherical harmonics $Y_l^m$.  Due to the axial symmetry around the axis connecting the centers of the two stars, $m$ is zero, and since the tidal deformation is dominantly quadrupolar, with no dipole, the leading contribution is at $l=2$ \cite{Hinderer:2009ca}. Hence the full perturbed metric $g_{\alpha\beta}+h_{\alpha\beta}$ (where $g_{\alpha\beta}$ is the metric from (\ref{sphericalmetric})) is written as
\beq
h_{\alpha\beta}={\rm diag}\left( e^{\nu(r)} H(r), e^{\mu(r)} H(r) ,r^2 K(r),r ^2 \sin^2 \theta K(r) \right) Y_2^0 (\theta ,\phi)~,
\eeq
where $e^{\mu(r)}$ and $e^{\nu(r)}$ are the functions in the solution to the unperturbed spherically symmetric metric~\eqref{sphericalmetric}:
\begin{align}
e^\mu(r) &=\left(1-2 G m(r)/r\right)^{-1}~, \\
\nu'(r) & = -\frac{2 p'(r)}{p(r) + \epsilon(r)}~,
\end{align}
and the Einstein equations relate the functions $K$ and  $H$:
\beq
K^\prime(r)= H^\prime(r) + H(r) \nu^\prime(r)~.
\eeq
Inserting the perturbed metric into Einstein's equations results in a second order differential equation for $H(r)$:
\beq
\begin{aligned}
H'' &= 2 H e^{\mu} \left\{ -2 \pi G\left[ 5 \epsilon + 9 p + \frac{d \epsilon}{dp} (\epsilon+p) \right] + \frac{3}{r^2} + 2 G^2e^{\mu} \left( \frac{m(r)}{r^2}+ 4 \pi r p \right)^2 \right\} \\
&\phantom{{}={}}+\frac{2}{r} H' e^{\mu} \left\{ -1 + \frac{Gm(r)}{r} + 2\pi Gr^2 (\epsilon - p) \right\}.
\end{aligned}
\eeq

To find solutions, one starts with a series expansion of $H$ very near the core of the star, at small $r$:
\beq
H(r)=a \,r^2+{\mathcal O}\!\left(r^4\right).
\eeq
The linear term drops out since the solution must be regular at $r=0$.
The size of the coefficient $a$ is linearly proportional to the size of the external perturbation, ${\mathcal E}_{ij}$, and is not an intrinsic property of the star, as is clear from the fact that it is simply a normalization coefficient for the solution to the linear ODE for $H$.  One can thus pick this coefficient arbitrarily in numerically solving for $H$.  The $l=2$ tidal Love number, on the other hand is an intrinsic property, and the value for $a$ drops out in calculating it. The value for $k_2$ can be calculated once $H$ is found, and matched at large $r$ onto the metric ansatz in Eq.~\eqref{asympmetric}.  It is given by
\beq
\begin{aligned}
k_2 &= \frac{8 C^5}{5} (1-2C)^2 [ 2 + 2 C (y-1) -y ] \\
&\phantom{{}={}}\times \left\{ 2 C [ 6- 3y + 3C (5y-8)]+4C^3[ 13-11 y + C (3y-2)+2C^2(1+y)] \right. \\
&\phantom{{}={}}+ \left. 3 (1-2C)^2 [ 2 - y + 2C (y-1)] \log (1-2C ) \right\}^{-1}~,
\end{aligned}
\eeq
where $C$ is the compactness parameter $GM/R$, and $y$ is obtained from the solution to $H$ evaluated on the surface of the star:
\beq
y = \frac{R H'(R)}{H(R)}~.
\eeq
Another dimensionless quantity, known as the dimensionless tidal deformability, is often found in the literature.  It is obtained from the definition of $k_2$ by factoring out the $C^5$ in front:
\beq
\bar{\lambda}=\frac{2k_2}{3C^5}=\frac{\lambda}{G^4M^5}~.
\label{eq:tidal-ligo}
\eeq

%%%%%%%%%%%%%%%%%%%%%%%%%%%%%%%%%%%%%%%%%%%%%%%%%%%%%%
%%%%%%%%%%%%%%%%%%%%%%%%%%%%%%%%%%%%%%%%%%%%%%%%%%%%%%
\section{Results and Fits}
\setcounter{equation}{0}
\setcounter{footnote}{0}
%%%%%%%%%%%%%%%%%%%%%%%%%%%%%%%%%%%%%%%%%%%%%%%%%%%%%%

We are now ready to present our results for the effects of adding a VE component to the innermost layer. We will use several different benchmark EoS's for modeling the NS's and investigate the consequences of the presence of VE in each of those cases. Two of the EoS's are more ``conservative" in the sense that the maximum stable NS mass that can be achieved just barely goes above $2 M_\odot$ (the maximum NS mass observed thus far is $M = (2.01 \pm 0.04) M_\odot$). The two more conservative models are the   AP4~\cite{Akmal:1998cf} and SLy~\cite{Douchin:2001sv} EoS's, which were also used as benchmarks by LIGO/Virgo~\cite{TheLIGOScientific:2017qsa}. We also consider the less restrictive model of Hebeler et al.~\cite{Hebeler:2013nza} which permits larger masses, up to nearly $3 M_\odot$. For the AP4 and SLy models we use the piecewise polytropic parametrization for all seven layers provided by Read et al.~\cite{Read:2008iy}. We have tabulated the corresponding parameters in Table~\ref{tab:parameters}. While the model of Hebeler et al.\ also uses a piecewise polytropic EoS for the innermost three layers, for the outer four layers corresponding to the crust they use a semi-analytic expression.  In their parametrization, the outer crust is modeled by the BPS EoS~\cite{Baym:1971pw} assuming $\beta$ equilibrium\footnote{$\beta$-equilibrium corresponds to equal rates of neutron decay and proton capture of electrons.}, followed by a layer for which chiral EFT (valid up to the nuclear saturation density around 0.18 baryons$/{\rm fm}^3$) is used to obtain the EoS. This semi-analytic expression is consistent with the piecewise polytropic approach of AP4 and SLy. 
\renewcommand{\arraystretch}{1.15}
\begin{table}[t]
	\begin{center}
		\begin{tabular}{| c | c | c | c |}
			\hline
			& \textbf{SLy} & \textbf{AP4} & \textbf{Hebeler} \\ 
			\hline
			$K_1$ & \multicolumn{2}{| c |}{$9.27637 \times 10^{-6}$} & See~\cite{Hebeler:2013nza}\\ \hline
			$p_1$ & \multicolumn{2}{| c |}{$(0.348867)^4$} & \\ 
			$p_2$ & \multicolumn{2}{| c |}{$(7.78544)^4$} & See~\cite{Hebeler:2013nza}\\ 
			$p_3$ & \multicolumn{2}{| c |}{$(10.5248)^4$} & \\ \hline
			$p_4$ & $(40.6446)^4$ & $(41.0810)^4$ & $(72.2274)^4$\\ 
			$p_5$ & $(103.804)^4$ & $(97.1544)^4$ & $(102.430)^4$\\ 
			$p_6$ & $(176.497)^4$ & $(179.161)^4$ & $(149.531)^4$\\ \hline
			$\gamma_1$ & \multicolumn{2}{| c |}{$1.58425$} & \\
			$\gamma_2$ & \multicolumn{2}{| c |}{$1.28733$} & \\
			$\gamma_3$ & \multicolumn{2}{| c |}{$0.62223$} & \\
			$\gamma_4$ & \multicolumn{2}{| c |}{$1.35692$} & \multirow{-4}{*}{See~\cite{Hebeler:2013nza}}\\ \hline
			$\gamma_5$ & $3.005$ & $2.830$ & $4.5$\\
			$\gamma_6$ & $2.988$ & $3.445$ & $5.5$\\
			$\gamma_7$ & $2.851$ & $3.348$ & $3$\\ \hline
		\end{tabular}		
	\caption{The parameters used for each EoS. The exponents $\gamma_i$ are dimensionless, the various pressures have units of $\textnormal{MeV}^4$, and $K_1$ is in units of $\textnormal{MeV}^{4-4\gamma_1}$. The Hebeler et al.\ parametrization~\cite{Hebeler:2013nza} uses a semi-analytic expression which is not piecewise polytropic in the outer region of the star, and thus cannot be displayed in the table.
	\label{tab:parameters}}
	\end{center}
\end{table}
 
Varying the EoS leads to more or less compact NS's, whose deformability will also change. The compactness of the NS can be characterized by the radius of a NS with a fixed mass. The deformability describes how much the NS reacts to the presence of the gravitational field of the second NS in the binary merger event and is characterized by the tidal deformability. 
  In the first subsection, we present our results for the mass versus radius, $M(R)$, curves of neutron stars with different nuclear EoS's including the effect of VE, while in the second, we study the tidal deformability 
and comment on the potential for LIGO/Virgo to discern between models with different assumptions about the change in VE in exotic phases of QCD. 

%%%%%%%%%%%%%%%%%%%%%%%%%%%%%%%%%%%%%%%%%%%%%%%%%%%%%%
\subsection{\texorpdfstring{$\bm{M(R)}$}{M(R)} Results}
We first present the results for the mass versus radius curves for the three different benchmark equations of state, whose parameters are displayed in Table~\ref{tab:parameters}.  These three benchmarks cover a realistic range of possible EoS's, with a wide variation in the maximum possible stable neutron star mass. We take care not to violate basic constraints on the behavior of high density QCD matter. For example, when pressures near the center of the star become very large, and relativistic effects dominate, one must ensure that the equation of state does not violate causality.  Causality requires that the speed of sound in the fluid does not exceed the speed of light:
\beq
v_\textnormal{s}=\sqrt{\frac{d p}{d \epsilon}} \le1~.
\eeq 
However, using simple  EoS models this condition is often violated for very large central pressures. Such violation does not imply the instability of the NS, but is rather an indication that the ansatz for the EoS is no longer a good approximation of the underlying nuclear physics in that region.  Such causality violation would never arise in the true QCD equation of state at very high densities.

The true stability condition for the central pressures that a neutron star can support is given by 
\beq
\frac{\partial M}{\partial p_{\text{center}}}>0~.
\label{eq:stability}
\eeq
This constraint arises from considerations of radial pulsation modes of the star, and is directly associated with stability of the fundamental mode of oscillation~\cite{bardeenthornemeltzer}.  The relation in Eq.~\eqref{eq:stability} above can be violated when the jump in the energy density is too large~\cite{Seidov71}:
\beq
\epsilon_+-\epsilon_- \ge \frac{1}{2}\epsilon_-+\frac{3}{2}p_6~.
\label{eq:ejump}
\eeq
Above this bound, the mass of the NS no longer increases with increasing core pressure, and the NS is unstable~\cite{Alvarez-Castillo:2017qki,Alford:2017qgh,Alford:2013aca,Seidov71,Shaeffer83}.

We note that the condition in Eq.~\eqref{eq:stability} can be satisfied for two stars of the same mass, but different internal pressures~\cite{Glendenning00}, corresponding to different phases in the core of the star.
In such cases, the critical energy density jump exceeds that in Eq.~\eqref{eq:ejump} at the transition.  However, even with this instability, one sometimes finds for $p>p_6$, that there is a disconnected class of solutions that does not exceed the bound in Eq.~\eqref{eq:stability}. The possibility then arises that some of the exotic, disconnected solutions have the same masses as some of the normal, lower pressure solutions.

Which of the two conditions, causality or monotonicity, will limit the central pressure depends on the EoS. For AP4 and SLy, the limit is set by causality. This bound can be avoided by modifying the EoS via a ``causal extension"~\cite{Hebeler:2013nza} into the regions where the pressure exceeds the maximal value set by the causality bound. For the models we are working with, we have found that this extension simply flattens out the curves at the point where causality is violated, and hence does not change the value of the maximum mass significantly. For this reason we have chosen not to make this modification and ended the curves at the point where the speed of sound reaches unity.

\begin{figure}[t]
	\centering
	\begin{subfigure}[b]{0.49\textwidth}
		\includegraphics[width=\textwidth]{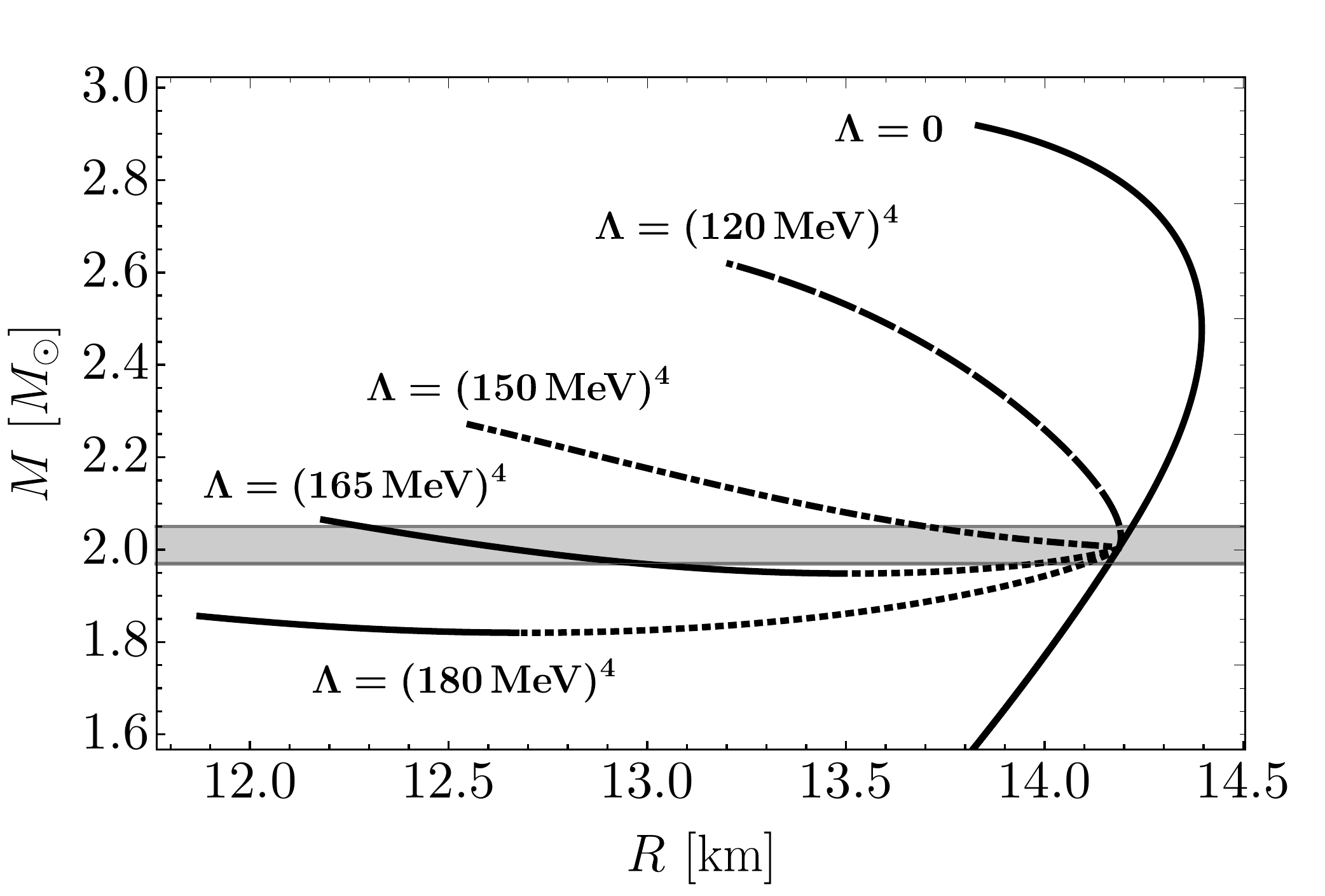}
	\end{subfigure}
	\begin{subfigure}[b]{0.49\textwidth}
		\includegraphics[width=\textwidth]{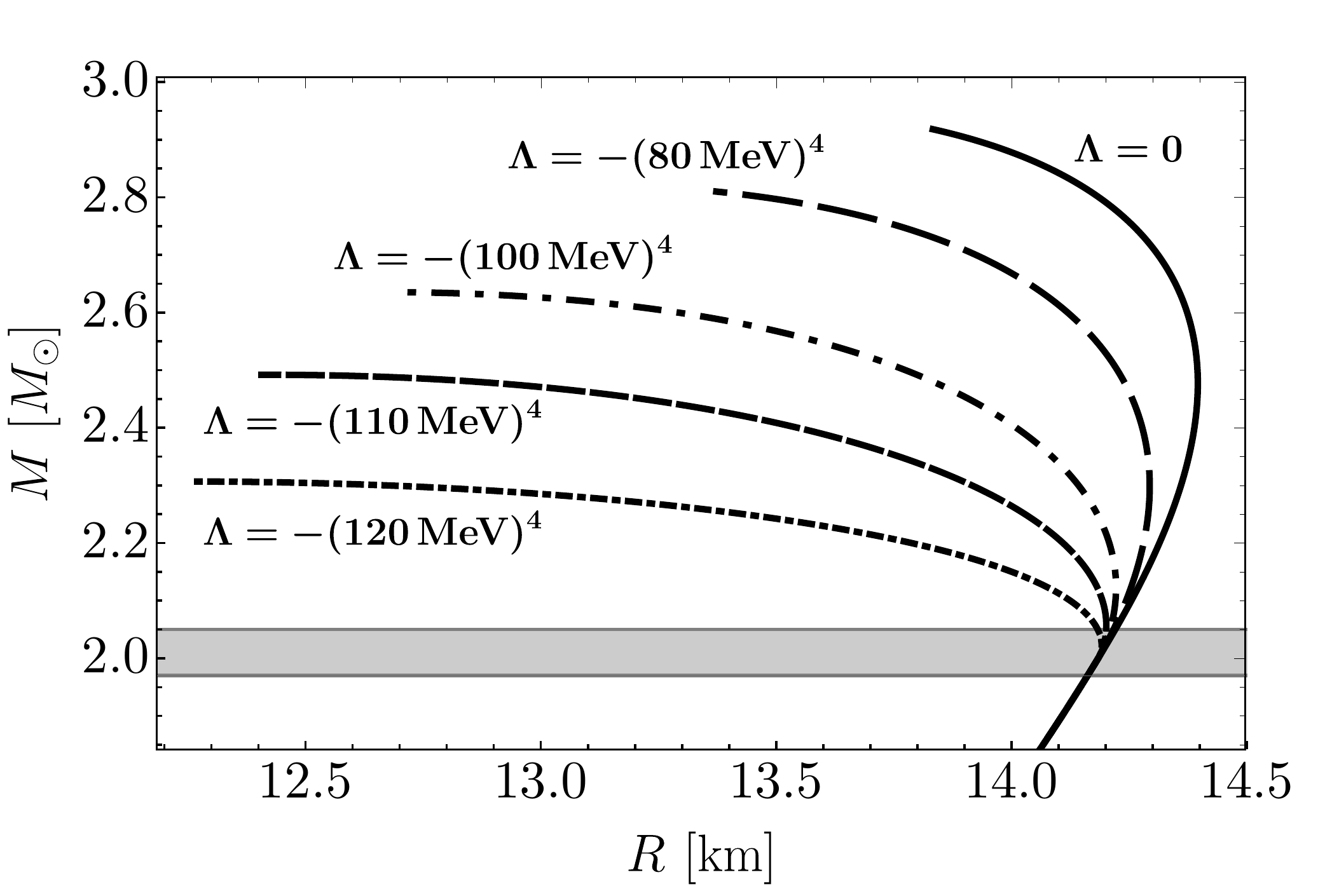}
	\end{subfigure}
	\caption{\small Mass versus radius curves corresponding to the stiff parametrization of Hebeler et al.~\cite{Hebeler:2013nza} with $\alpha=3$. Dotted curves in the plot on the left correspond to unstable configurations violating Eq.~\eqref{eq:stability}.  Positive values of $\Lambda$ are shown in the plot on the left, and negative ones on the right.}
	\label{fig:HebelerMR}
\end{figure}

The $M(R)$ curve and the effect of VE for the Hebeler et al.\ EoS~\cite{Hebeler:2013nza} are shown in Fig.~\ref{fig:HebelerMR}.  Each curve is obtained by varying the pressure at the center of the star but keeping all of the other parameters fixed. We have fixed $\alpha =3$ in this plot, as well as all those that follow.\footnote{Taking $\alpha$ to be small reduces the change in the curves relative the $\Lambda = 0$ case, however small values of $\alpha$ are not representative of most phase transitions, which are typically accompanied by a change in the energy density as well as the vacuum energy.}  When the central pressure is greater than $p_6$, the value of $\Lambda$ becomes relevant and the other curves depart from the behavior of the $\Lambda=0$ case. Dotted parts of the curves correspond to unstable regions, i.e.\ solutions of the TOV equations in which the stability condition \eqref{eq:stability} is violated. The shaded region represents the most massive neutron star measured to date, with a mass of $(2.01\pm0.04)M_\odot$~\cite{Demorest:2010bx}. Notice that for some positive values of $\Lambda$, i.e.\ when the jump in energy density is big enough according to Eq.~\eqref{eq:ejump}, we find a second stable region which is disconnected from the main branch, as discussed above.  This means that for a given mass, there are two possible types of stars, one with no exotic phase in the core, and another with a significant portion of the star in the new phase.   This gives rise to interesting effects, both for $M(R)$ curves and in GW observables. For example, assuming that the $\Lambda=(165\MeV)^4$ curve is the correct one, it would be possible to observe two $2M_\odot$ neutron stars with significantly different radii.  That is, there are two stable configurations for stars with the same mass.   It is quite interesting that the physics of QCD may allow for a plethora of different compact objects, with population numbers depending on the conditions of their formation.  

Our procedure for introducing the VE for this model is the following. In order to make sure that all values of $\Lambda$ considered here are compatible with a neutron star mass of $(2.01\pm0.04)M_\odot$, we stop the next-to-innermost polytropic region as soon as the mass reaches $2.00M_\odot$. This corresponds to choosing a critical pressure $p_6\approx(150\MeV)^4$. Once the critical pressure is reached, we transition into the innermost polytropic region with a nonzero VE,  and we allow for the central pressure to be as high as possible without violating the causality bound.

\begin{figure}[t]
	\centering
	\begin{subfigure}[b]{0.48\textwidth}
		\includegraphics[width=\textwidth]{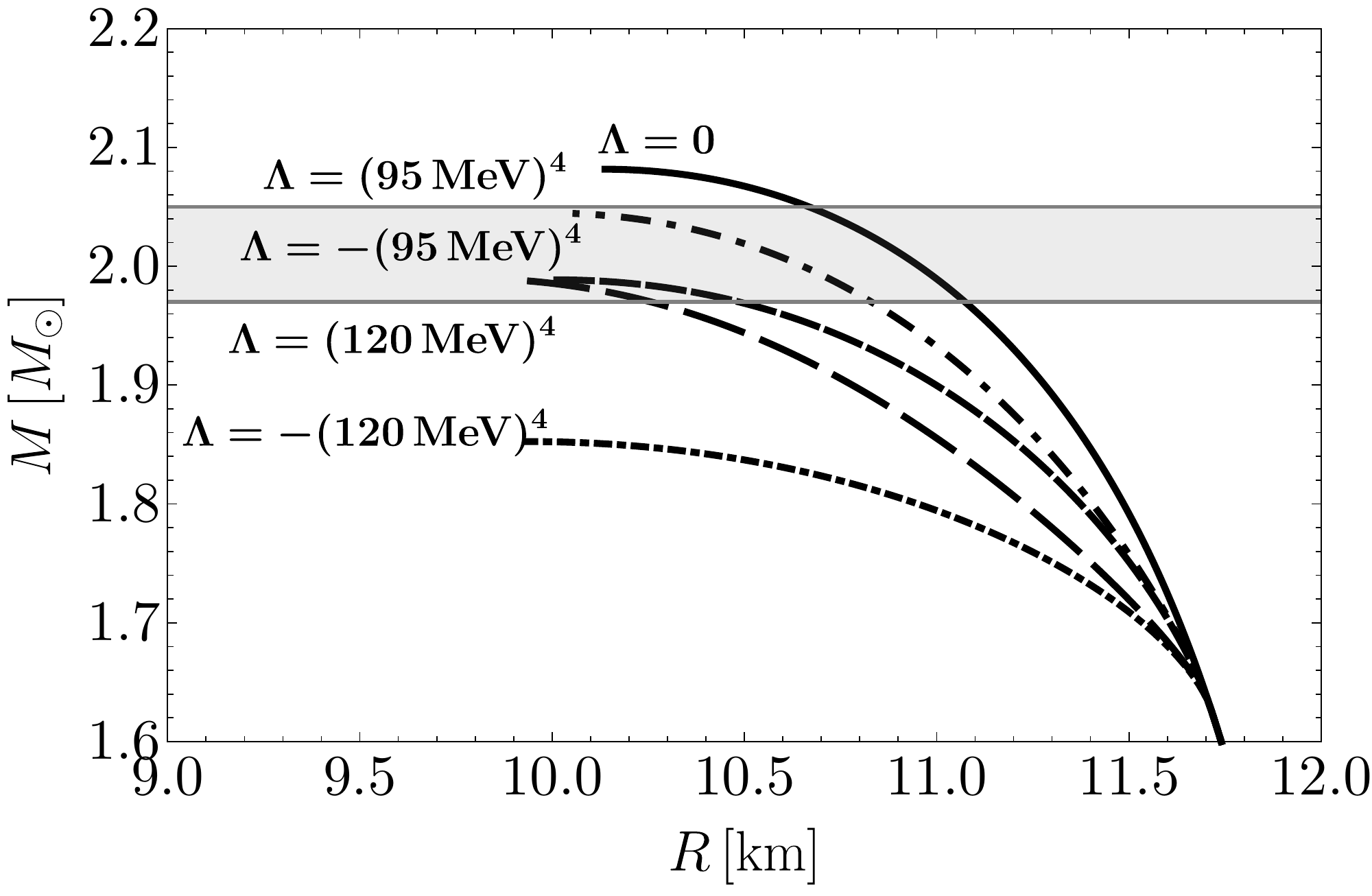}
		\caption{SLy EoS}
		\label{fig:sly-mr}
	\end{subfigure}
	~ %add desired spacing between images, e. g. ~, \quad, \qquad, \hfill etc. 
	\begin{subfigure}[b]{0.48\textwidth}
		\includegraphics[width=\textwidth]{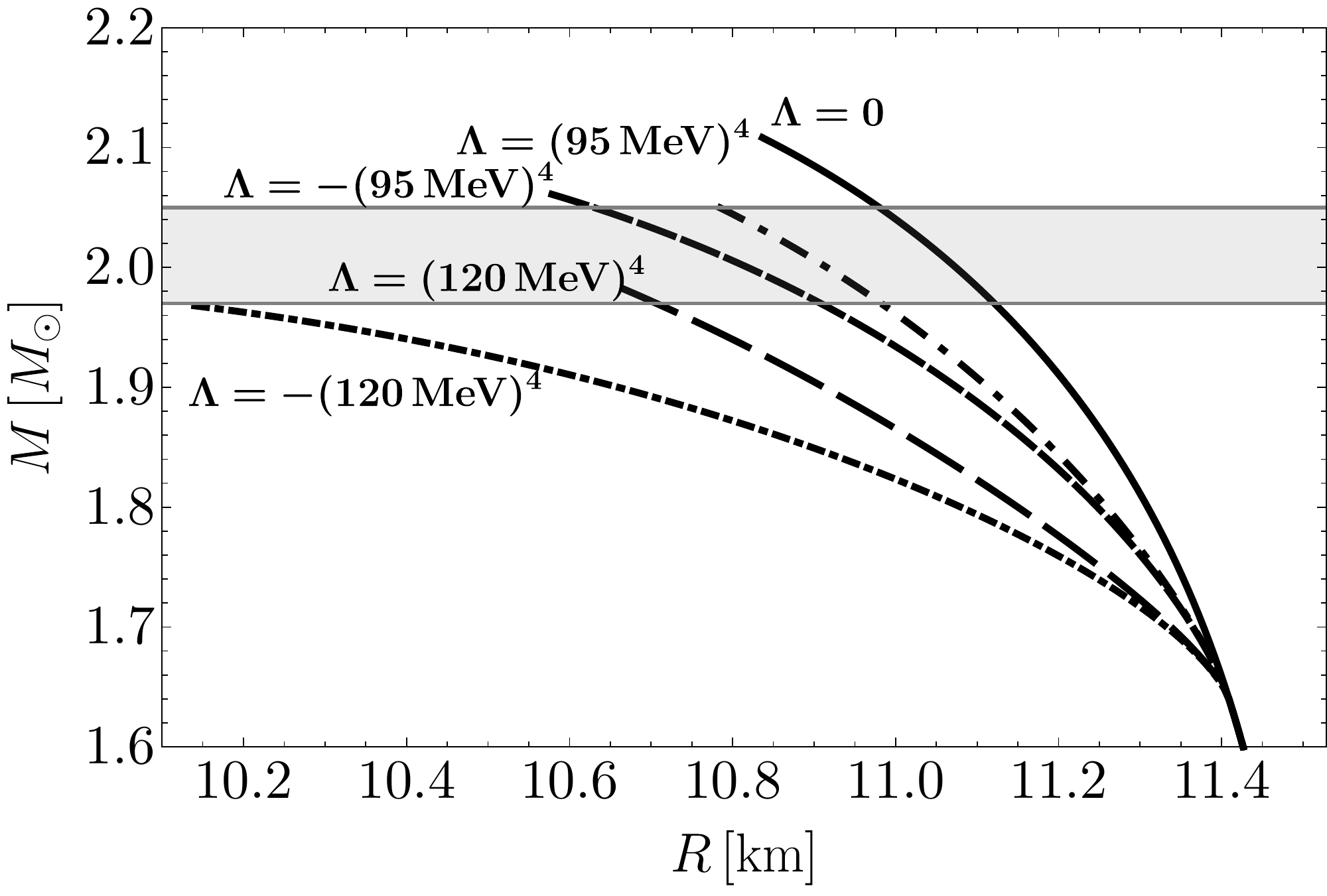}
		\caption{AP4 EoS}
		\label{fig:ap4-m3}
	\end{subfigure}
	\caption{\small $M(R)$ curves for the SLy and AP4 equations of state for various $\Lambda$ values on the seventh layer. For all the curves, the proportionality constant $\alpha$ in the jump equation \eqref{jump} is chosen to be $\alpha=3$. The gray region shows the allowed mass range of the heaviest neutron star, with mass $(2.01 \pm 0.04)M_{\odot}$.}
	\label{fig:mr-ap4-sly-plot}
\end{figure}

Next we present results for the AP4~\cite{Akmal:1998cf} and SLy~\cite{Douchin:2001sv} EoS models.     The $M(R)$ curves for the AP4 and SLy models with different values of the VE in the innermost layer are shown in Fig.~\ref{fig:mr-ap4-sly-plot}. One can again see that up to a certain critical mass, the curves corresponding to different $\Lambda$'s in the innermost layer do coincide with each other. The reason for this is that below this mass the central pressure is not high enough to enter the exotic high density phase of QCD. The critical pressure for the phase transition to occur is set by $p_6$ which is an input of the model. For the AP4 and SLy models, $p_6 \approx (179\MeV)^4$ and $p_6 \approx (176\MeV)^4$ respectively which correspond to a critical mass of approximately $1.6M_{\odot}$ for both models.

The plots for all three EoS's show that the maximal mass of the neutron star decreases for both positive and negative values of VE. This is a generic feature of neutron star models with phase transitions with vacuum energy in our study, and is due to the jump in the energy density across the phase transition.  This conclusion is similar to that obtained in previous works that study phase transitions without vacuum energy~\cite{Heiselberg:1999fe}.

%%%%%%%%%%%%%%%%%%%%%%%%%%%%%%%%%%%%%%%%%%%%%%%%%%%%%%

\subsection{Tidal Deformabilities and LIGO/Virgo}

Let us now discuss NS observables from GW's emitted during the merger of NS's. The  frequency versus time behavior of the waveform of the emitted gravitational wave, usually expressed in terms of the ``gravitational wave phase" appearing in the Fourier transform of the chirp, can be determined by an expansion in relativistic effects, starting at Newtonian order, and proceeding to post-Newtonian corrections in the velocity.  At dominant Newtonian order, where the two NS's are approximated by point masses, the waveform depends only on a particular combination of the masses called the chirp mass:
\beq
\mathcal{M}= \mu^{3/5} M_\text{tot}^{2/5} = \frac{(M_1 M_2)^{3/5}}{(M_1 + M_2)^{1/5}}~,
\eeq
where $\mu$ is the reduced mass of the system~\cite{Cutler:1994ys}.
For the recently observed merger event, GW170817, the chirp mass was measured to be $\mathcal{M}=1.188^{+0.004}_{-0.002} M_{\odot}$. 

Since this is the dominant contribution to the waveform, the chirp mass can be determined quite accurately.  However, the individual masses must be extracted from higher order velocity corrections to the waveform, and are thus more difficult to constrain.  At higher order, spin-couplings are important as well, and without information about the stars' rotational speeds and axes, precise extraction of the masses is impossible.  This information is in principle retrievable from measurements of the waveform, but is difficult as it relies on data near the end of the inspiral, where current experiments lose sensitivity, and where full numerical simulation of the merger event may be necessary~\cite{Dietrich:2017aum}.

 At present, the individual masses can only be estimated by using the chirp mass and some assumptions for the angular rotation frequency of the stars. For  GW170817 in the low-spin case, the estimated mass range is $1.36$--$1.60 M_{\odot}$ for the heavy star and $1.17$--$1.36 M_{\odot}$ for the light star, while for the high-spin case, there is considerably more possible variation in the masses: $1.36$--$2.26 M_{\odot}$ for the heavy star and $0.86$--$1.36 M_{\odot}$ for its less massive partner.

Similarly, it is not yet possible to measure individual tidal deformabilities. However, it is possible to constrain a weighted combination of the individual masses and deformabilities through their contribution to the gravitational wave phase at order $v^5$.  This so-called ``combined dimensionless tidal deformability" is defined as
\beq
\tilde{\Lambda}=\frac{16}{13}\frac{(M_1+12M_2)M_1^4\bar{\lambda}_1+(M_2+12M_1)M_2^4\bar{\lambda}_2}{(M_1+M_2)^5}~.
\label{eq:lambdatilde}
\eeq
For the recent event GW170817, the current constraint placed on $\tilde{\Lambda}$ is $\leq 800$ for the low-spin assumption and $\leq 700$ for the high-spin case.  In the low-spin case, the neutron star masses are probably too low to contain an exotic QCD phase, and thus event GW170817 would not contain information about VE.  Of course, this may not be the case for future merger events, which may involve heavier NS's.  In the high-spin case, however, the inner core could be in the exotic phase, and the constraints from GW170817 are relevant for studying VE.

\begin{figure}[t]
	\centering
	\begin{subfigure}[b]{0.49\textwidth}
		\includegraphics[width=\textwidth]{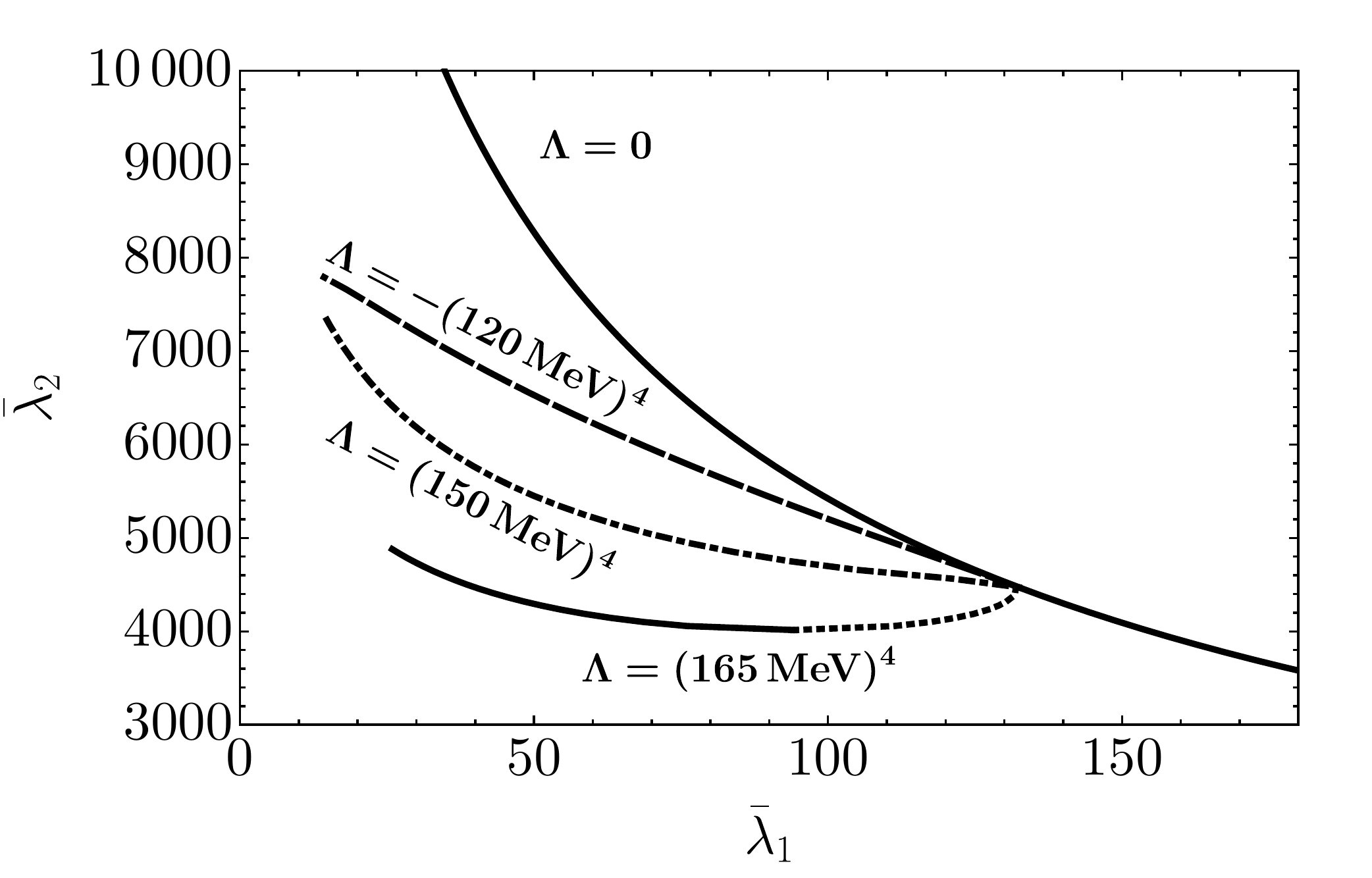}
		\caption{\small $\mathcal{M}=2^{-1/5}1.4M_\odot\approx1.22M_\odot$}
	\end{subfigure}
	\begin{subfigure}[b]{0.49\textwidth}
		\includegraphics[width=\textwidth]{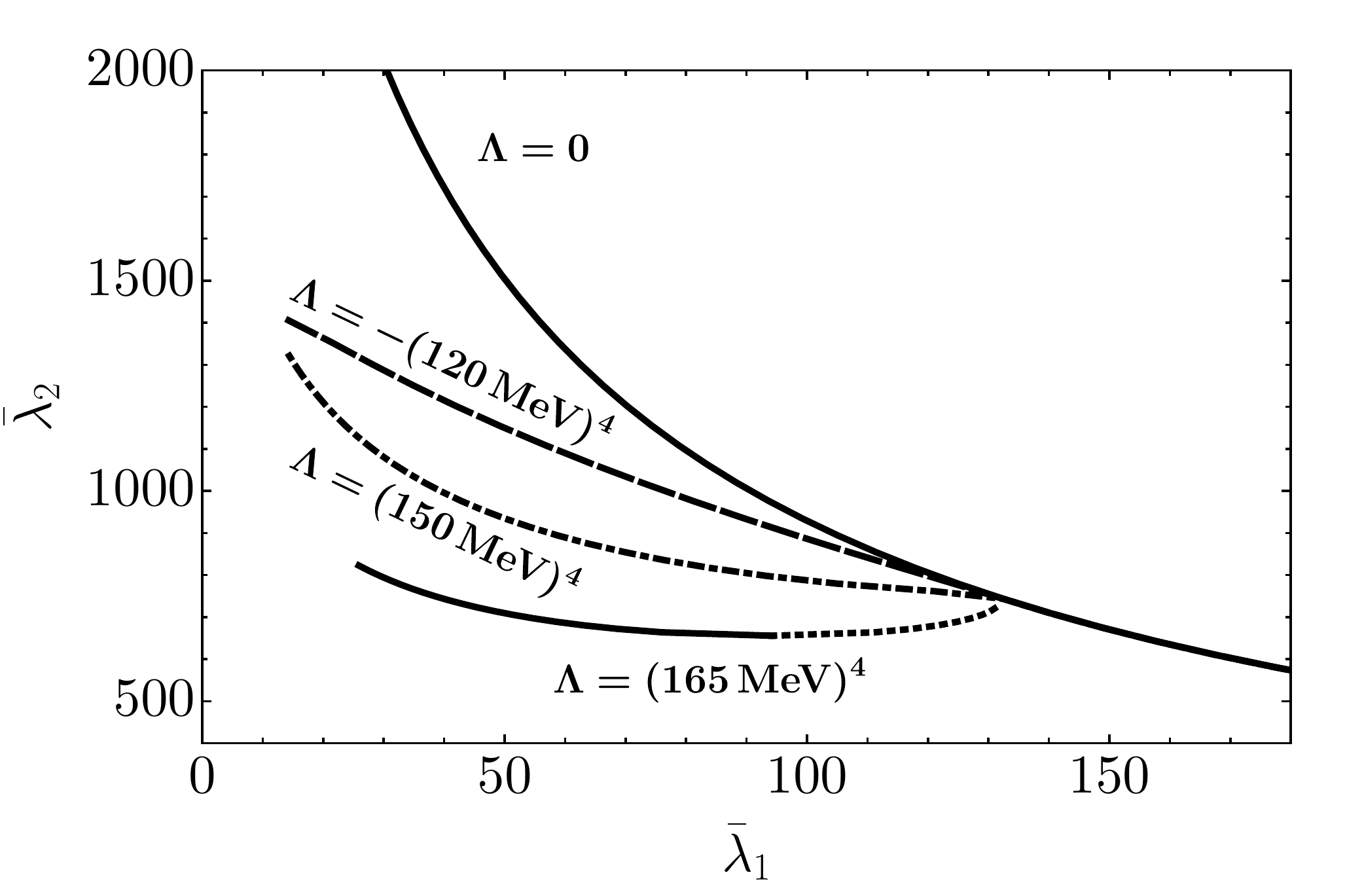}
		\caption{\small $\mathcal{M}=2^{-1/5}1.7M_\odot\approx1.48M_\odot$}
	\end{subfigure}
	
	\begin{subfigure}[b]{0.49\textwidth}
		\includegraphics[width=\textwidth]{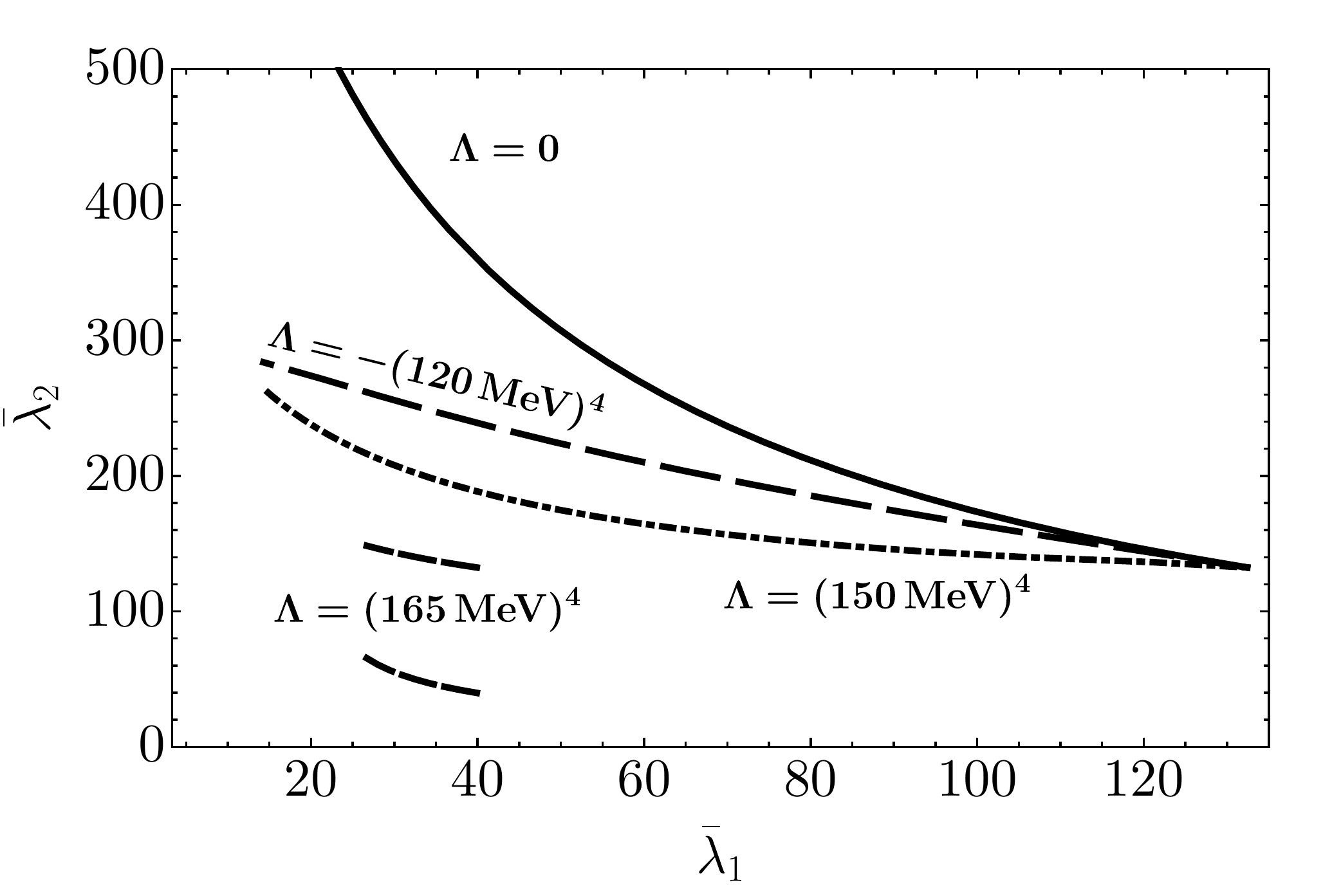}
		\caption{\small $\mathcal{M}=2^{-1/5}2.0M_\odot\approx1.74M_\odot$}
	\end{subfigure}	
	\caption{\small Tidal deformabilities for the Hebeler et al.\ parametrization with $\alpha=3$. Each plot corresponds to a different chirp mass. Dotted parts of the curves with $\Lambda = (165\MeV)^4$ correspond to unstable configurations. In all cases, the deviation from the $\Lambda=0$ curve is significant.} 
	\label{fig:HebelerTidal}
\end{figure}

\begin{figure}
	\centering
	\begin{subfigure}[b]{0.49\textwidth}
		\includegraphics[width=\textwidth]{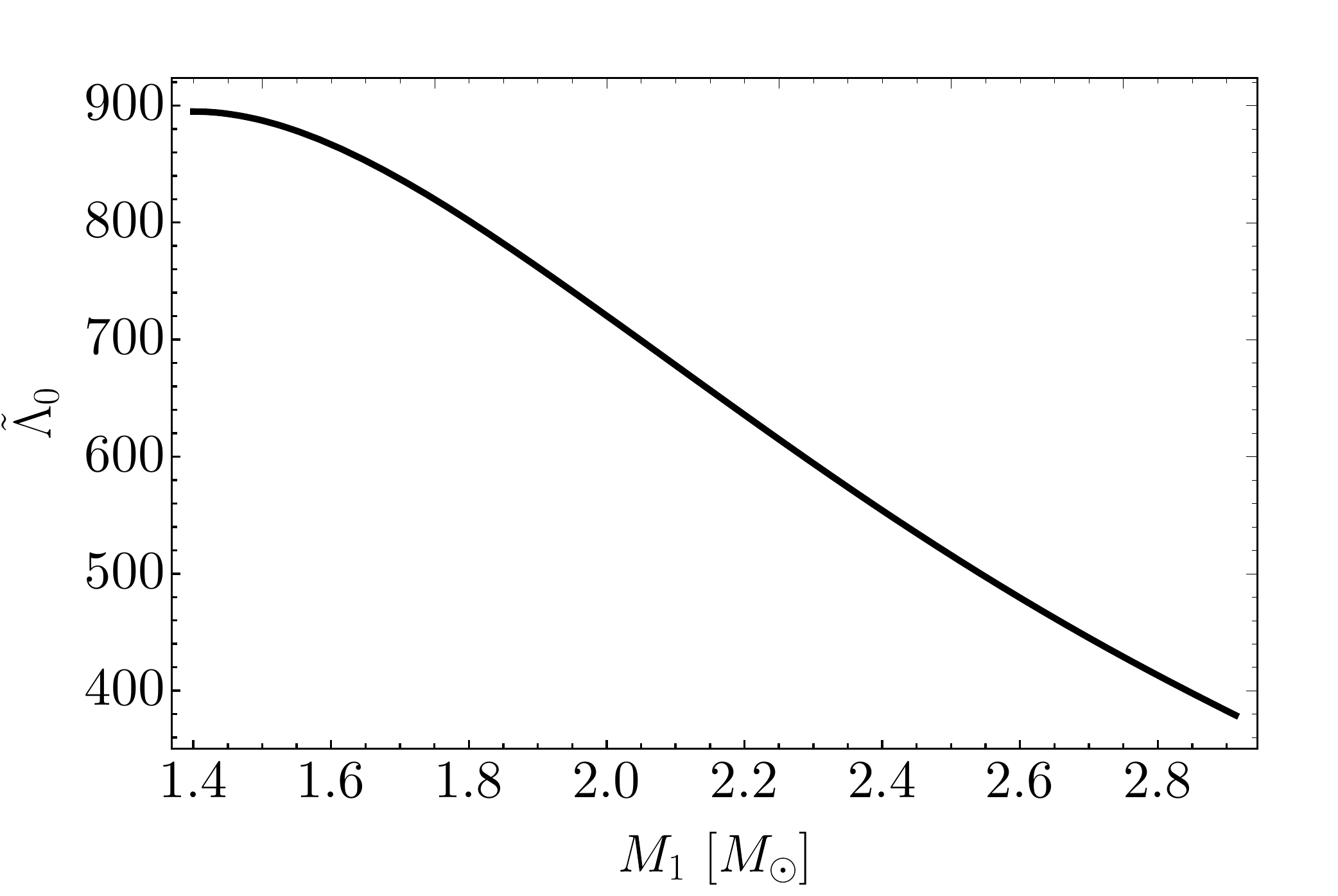}
		\caption{\small $\mathcal{M}=2^{-1/5}1.4M_\odot\approx1.22M_\odot$}
	\end{subfigure}
	\begin{subfigure}[b]{0.49\textwidth}
		\includegraphics[width=\textwidth]{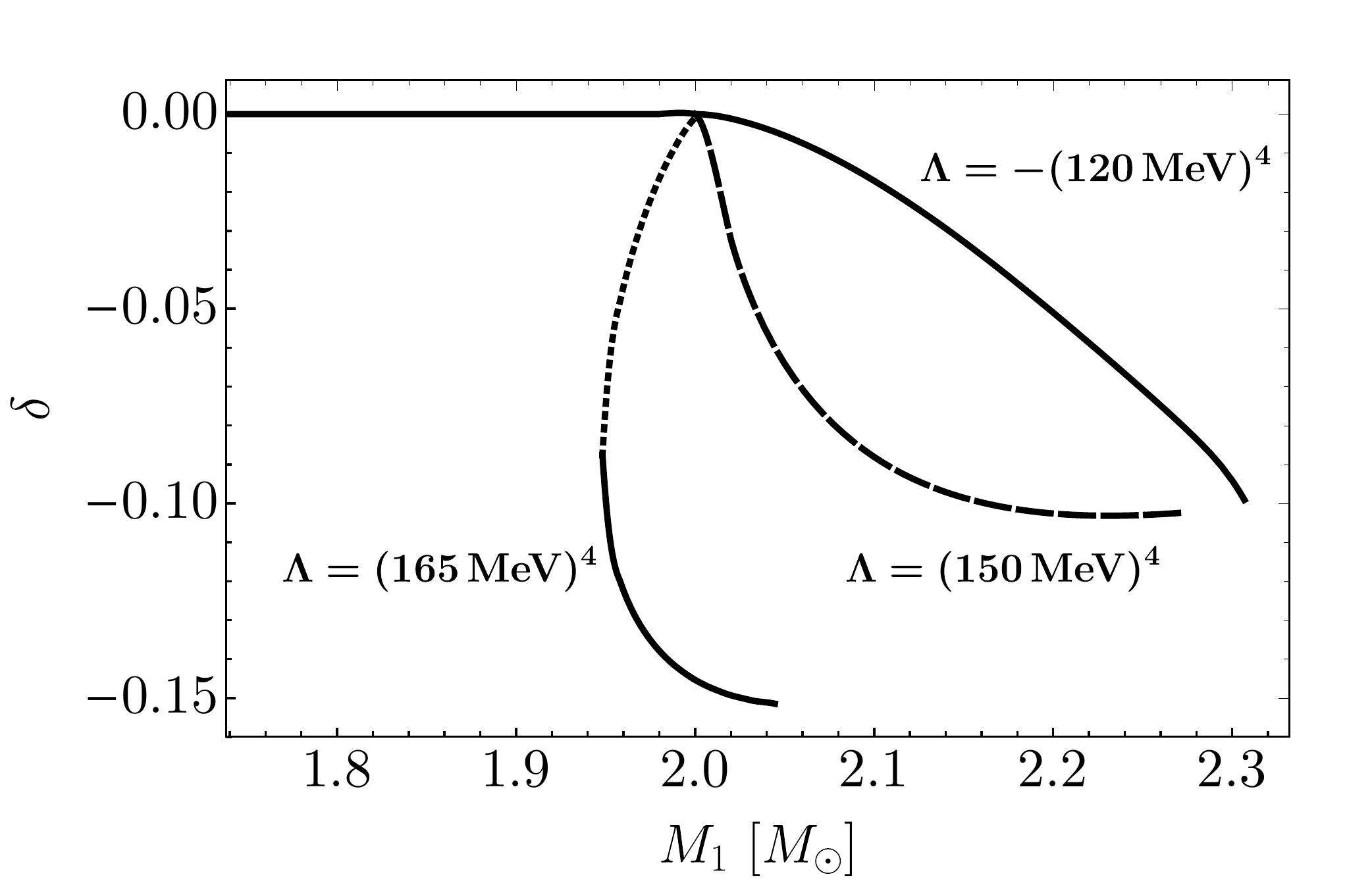}
		\caption{\small $\mathcal{M}=2^{-1/5}1.4M_\odot\approx1.22M_\odot$}
	\end{subfigure}
	
	\begin{subfigure}[b]{0.49\textwidth}
		\includegraphics[width=\textwidth]{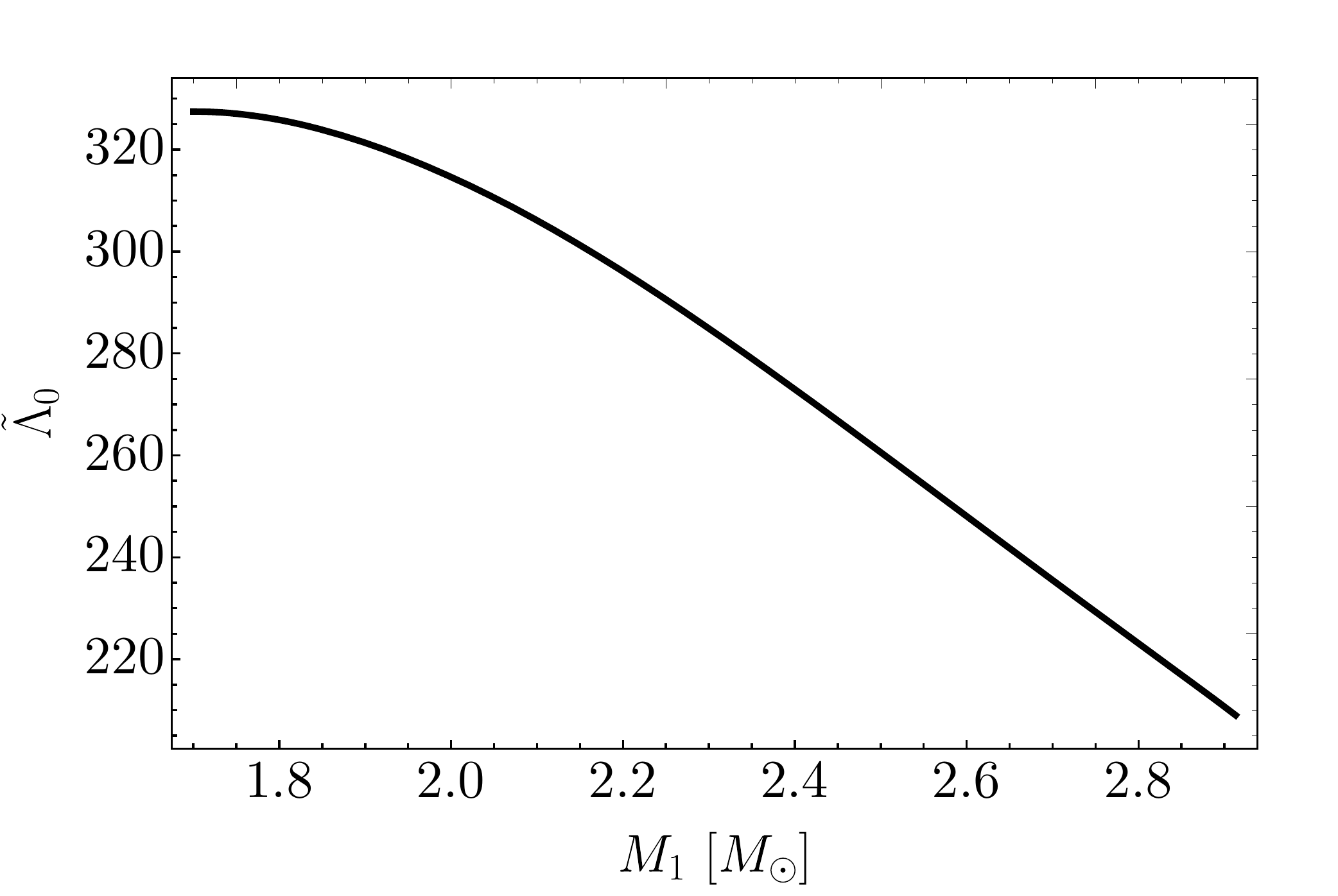}
		\caption{\small $\mathcal{M}=2^{-1/5}1.7M_\odot\approx1.48M_\odot$}
	\end{subfigure}
	\begin{subfigure}[b]{0.49\textwidth}
		\includegraphics[width=\textwidth]{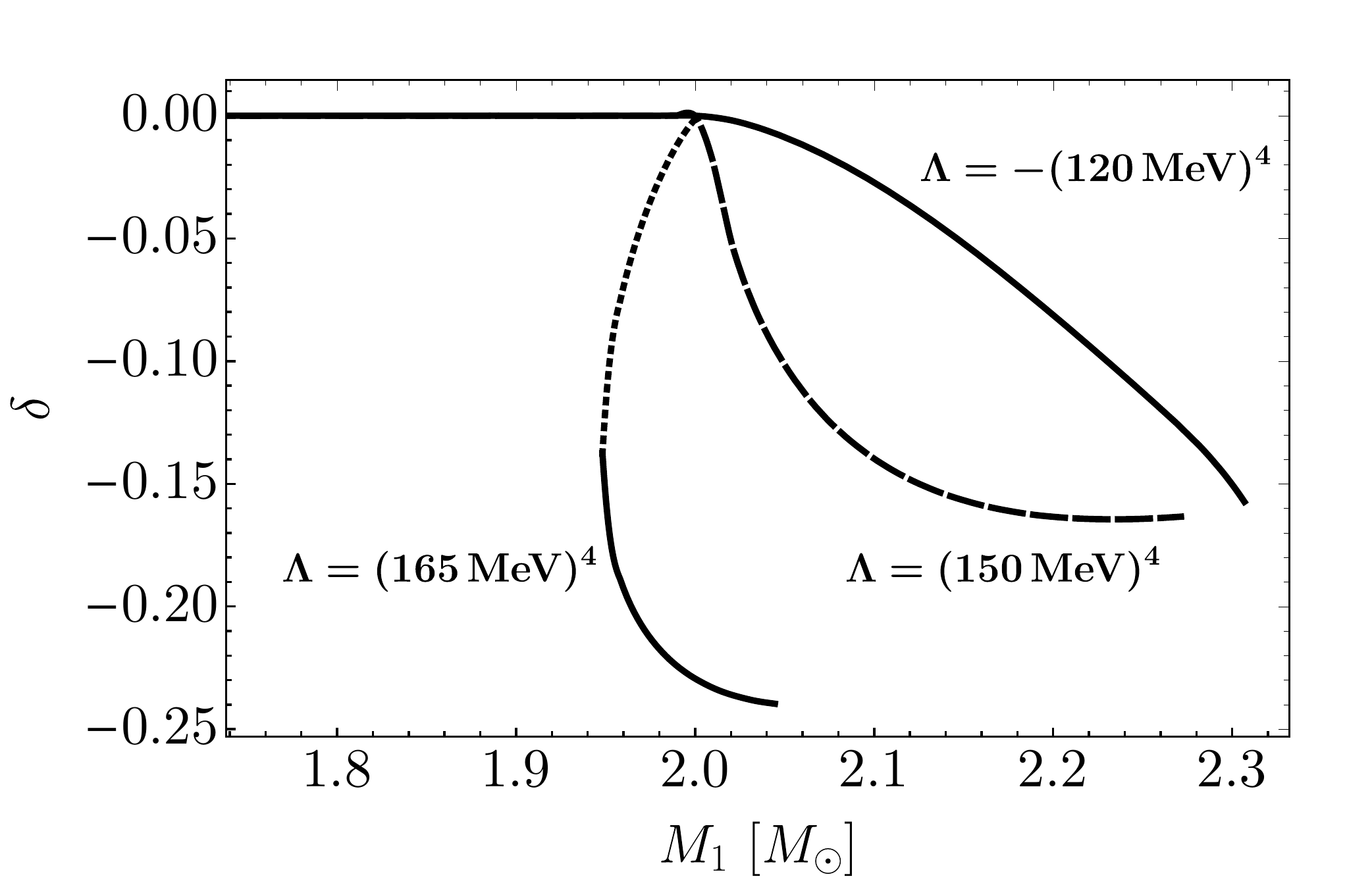}
		\caption{\small $\mathcal{M}=2^{-1/5}1.7M_\odot\approx1.48M_\odot$}
	\end{subfigure}
	
	\begin{subfigure}[b]{0.49\textwidth}
		\includegraphics[width=\textwidth]{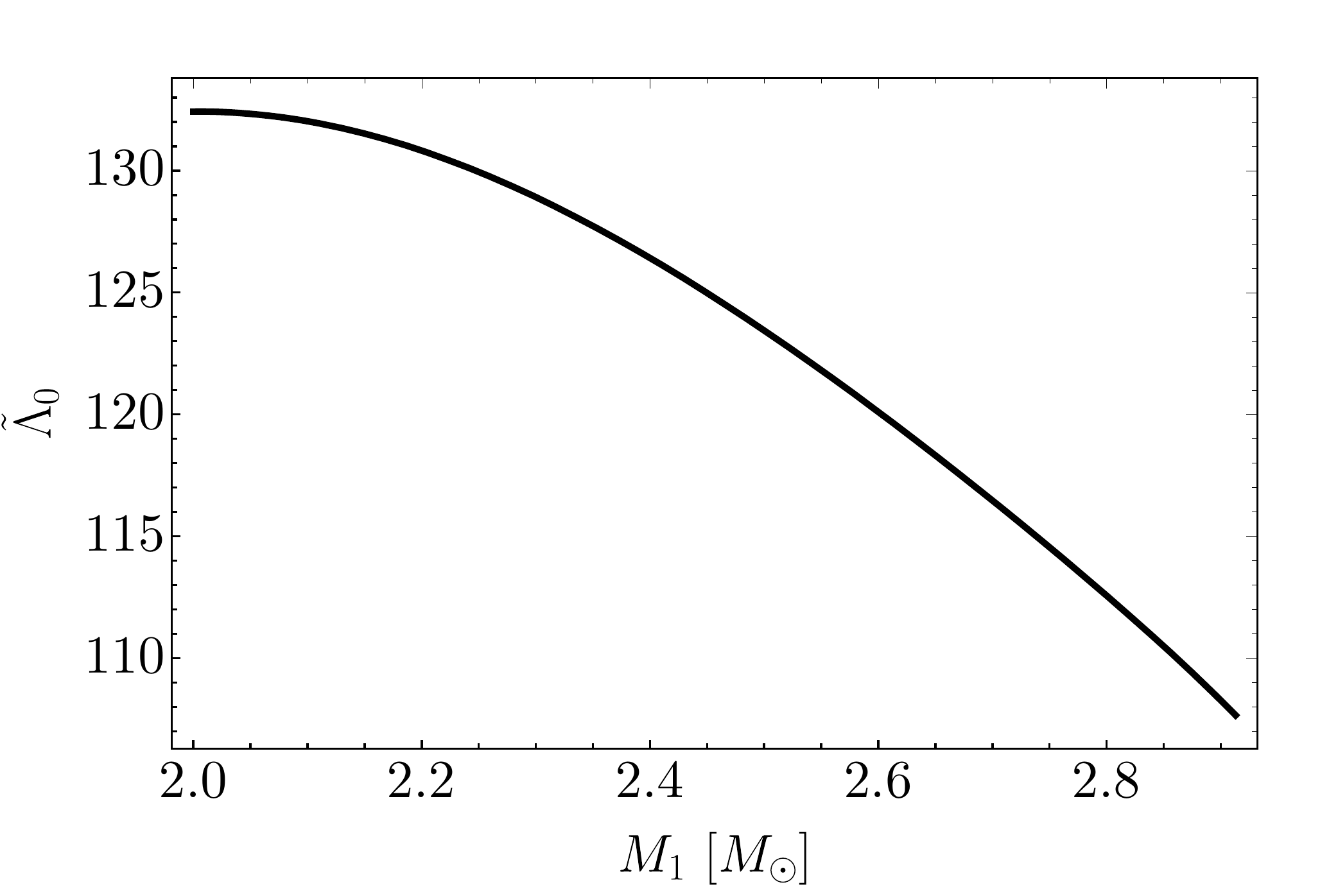}
		\caption{\small $\mathcal{M}=2^{-1/5}2.0M_\odot\approx1.74M_\odot$}
	\end{subfigure}	
	\begin{subfigure}[b]{0.49\textwidth}
		\includegraphics[width=\textwidth]{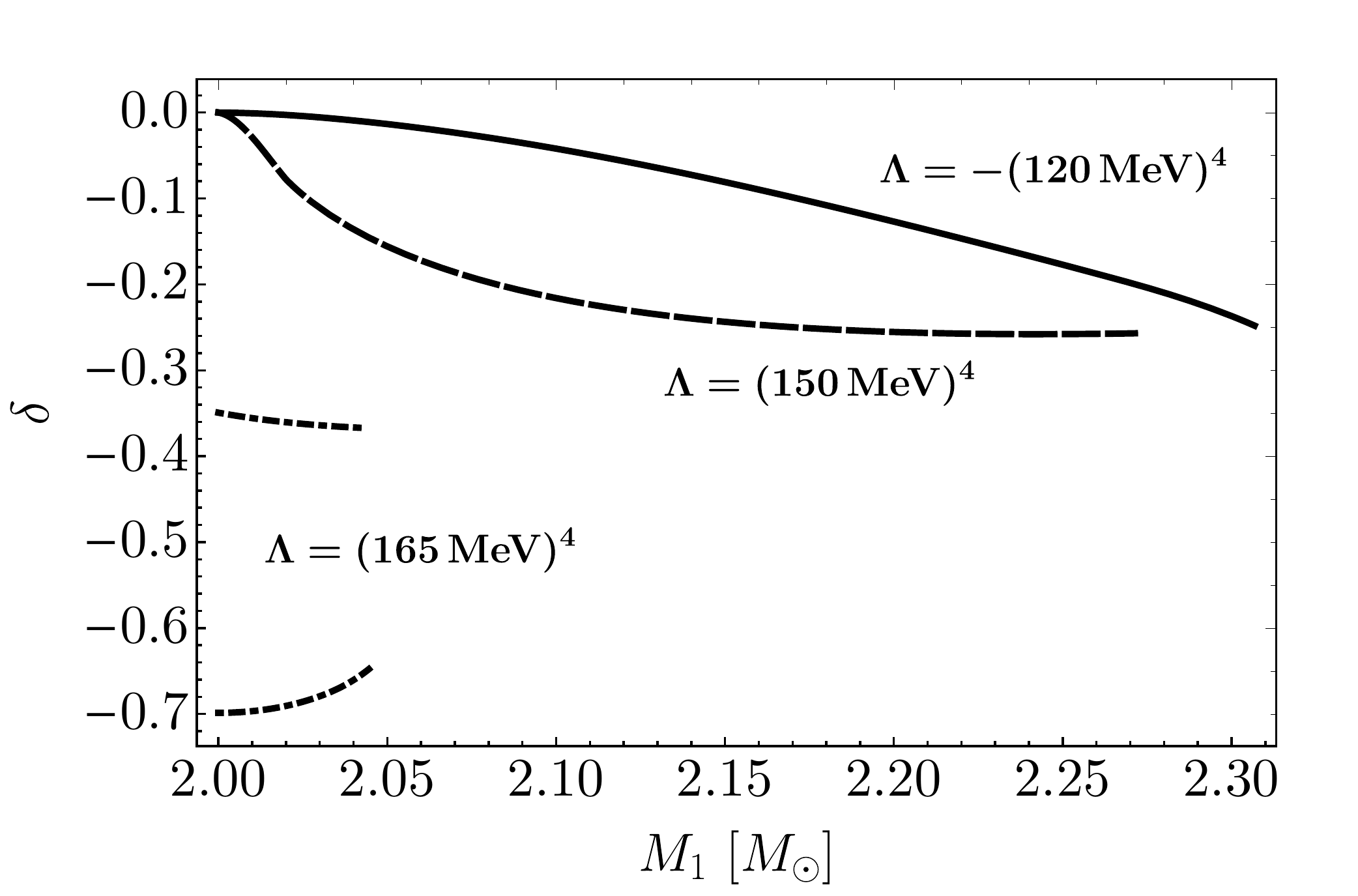}
		\caption{\small $\mathcal{M}=2^{-1/5}2.0M_\odot\approx1.74M_\odot$}
	\end{subfigure}	
		\caption{\small Plots on the right show the relative deviation of the combined dimensionless tidal deformability, $\tilde{\Lambda}$, as a function of the heaviest star mass for the Hebeler et al.\ parametrization with $\alpha=3$ for various values of the chirp mass.   Plots on the left show $\tilde{\Lambda}$ for vanishing VE for the same chirp masses.  Dotted parts of the curves correspond to unstable configurations. The disconnected branches associated with two stable NS configurations allow for the largest deviations.}
	\label{fig:HebelerGWphase}
\end{figure}

\begin{figure}[t]
	\centering
		\begin{subfigure}[b]{0.48\textwidth}
			\includegraphics[width=\textwidth]{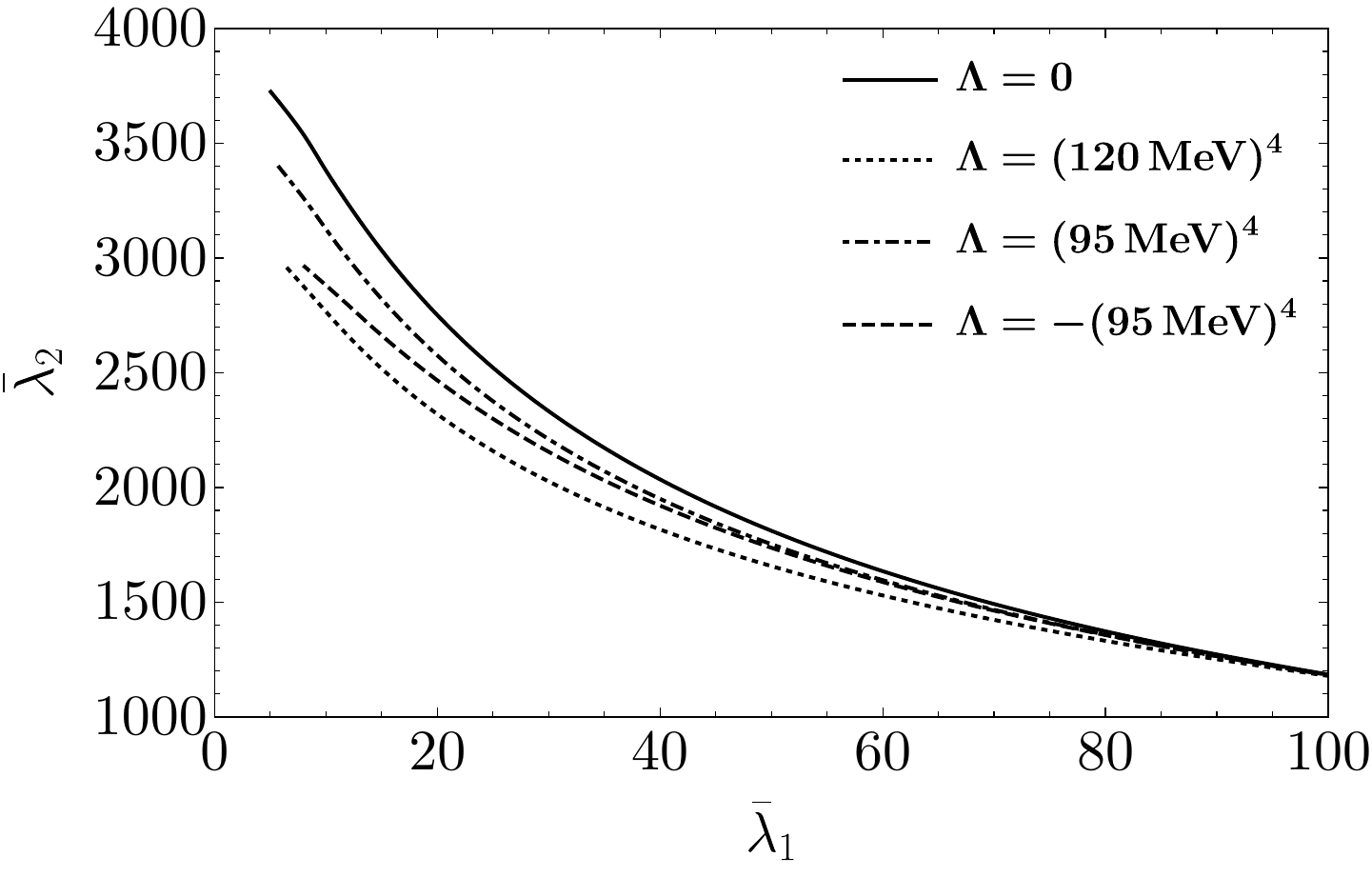}
			\caption{SLy EoS}
			\label{fig:sly-0CC-tidal}
		\end{subfigure}
		~ %add desired spacing between images, e. g. ~, \quad, \qquad, \hfill etc. 
		\begin{subfigure}[b]{0.48\textwidth}
			\includegraphics[width=\textwidth]{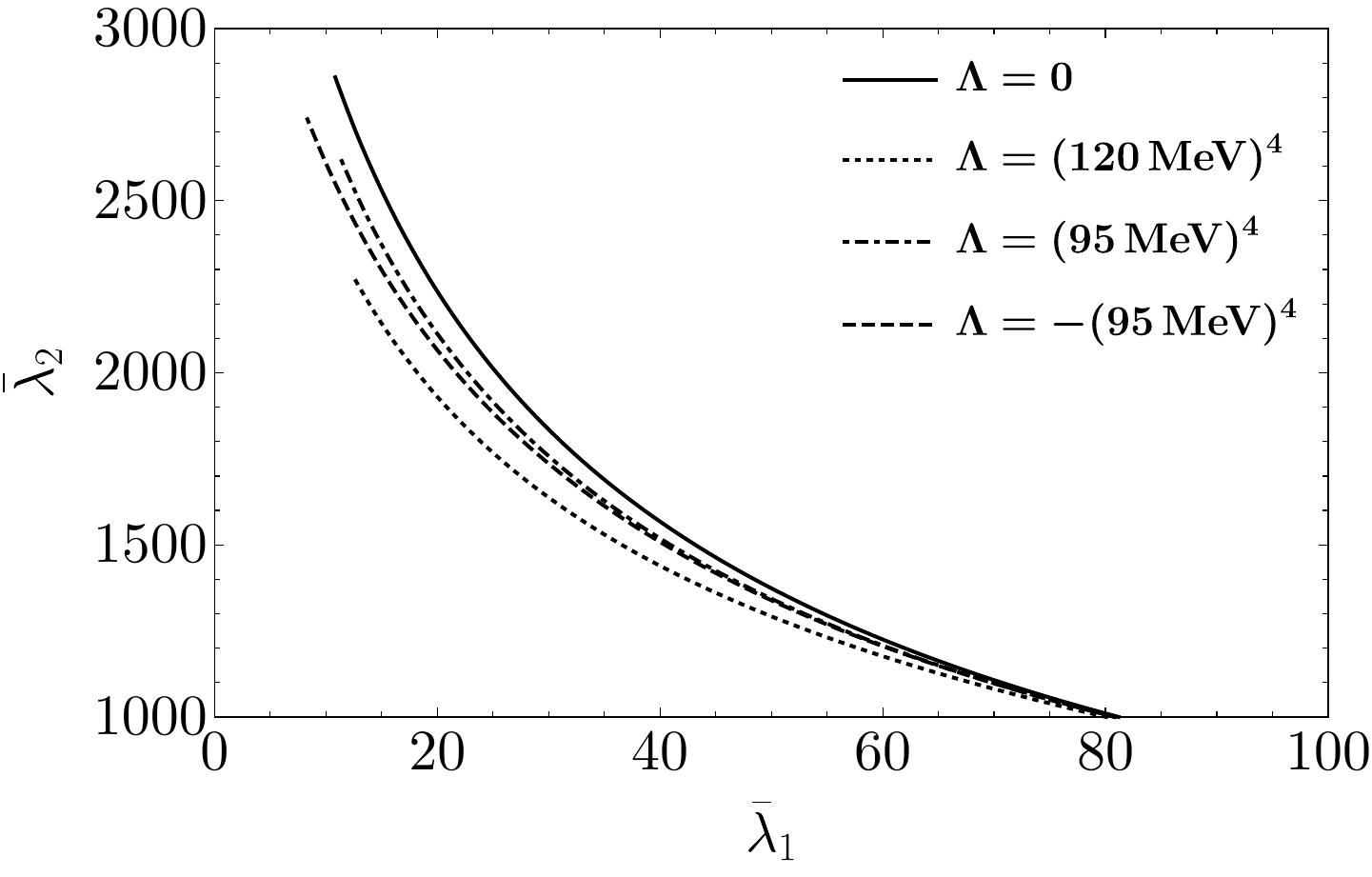}
			\caption{AP4 EoS}
			\label{fig:ap4-0CC-tidal}
		\end{subfigure}
	\caption{\small Tidal deformability curves for a neutron star binary with SLy and AP4  EoS's. The chirp mass is taken to be  $\mathcal{M}=1.188M_{\odot}$, which is the same value as in GW170817. $\bar{\lambda}_1$ and $\bar{\lambda}_2$ correspond to the dimensionless tidal deformability parameters for the heavy and light stars, respectively. Each curve is obtained by  varying the heavy star mass while holding the chirp mass fixed. The $\alpha$-parameter of \eqref{jump} is chosen to be $\alpha=3$. }
	\label{tidaldeform_ligo_slyap4}
\end{figure}

\begin{figure}[t]
	\centering
	\begin{subfigure}[b]{0.98\textwidth}
		\includegraphics[width=\textwidth]{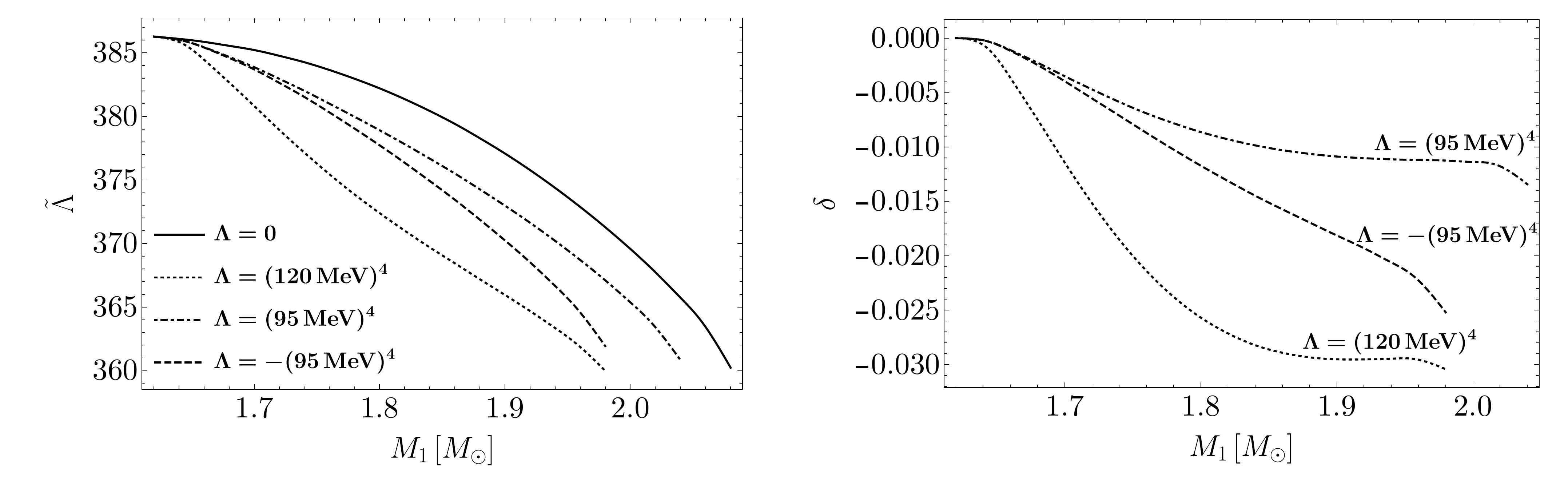}
		\caption{$\mathcal{M}=1.188M_{\odot}$}
	\end{subfigure}
	~ %add desired spacing between images, e. g. ~, \quad, \qquad, \hfill etc. 

	\begin{subfigure}[b]{0.98\textwidth}
		\includegraphics[width=\textwidth]{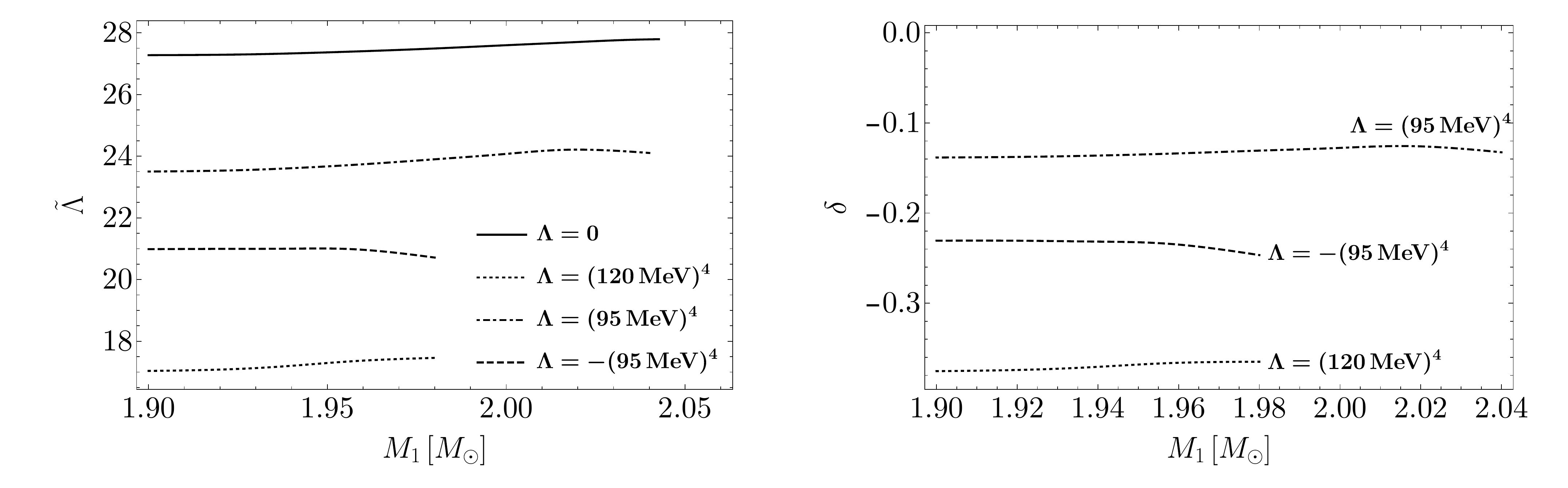}
		\caption{$\mathcal{M}=1.65M_{\odot}$}
	\end{subfigure}
	\caption{\small  Plot of the deviation of the combined dimensionless tidal deformability as a function of the heavy star mass for the SLy EoS with different values for the chirp mass. $\mathcal{M}=1.188M_{\odot}$ is the same as the one of GW170817, while $\mathcal{M}=1.65M_{\odot}$ corresponds to a chirp mass where if the two NS masses are equal they have a mass of $1.9M_{\odot}$. For the smaller chirp mass the effect is rather small, however for a higher chirp mass the effect can be as large as $38\%$.  The $\alpha$-parameter of \eqref{jump} is again chosen to be $\alpha=3$.}
	\label{fig:sly-dev}
\end{figure}

\begin{figure}[t]
	\centering
	\begin{subfigure}[b]{0.98\textwidth}
		\includegraphics[width=\textwidth]{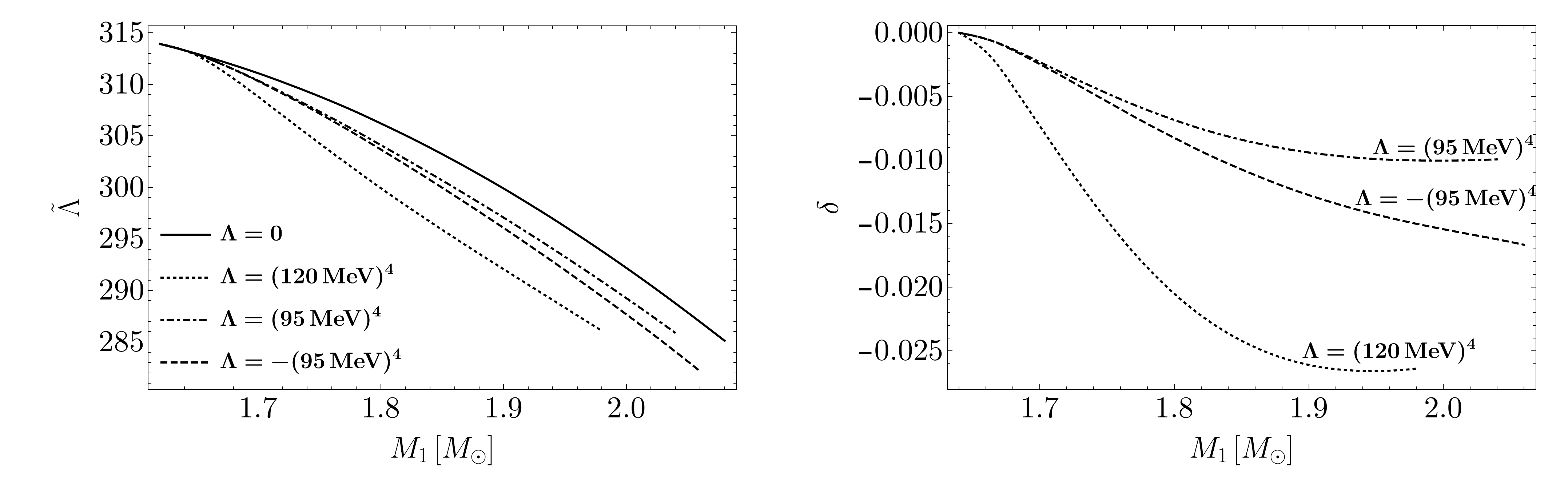}
		\caption{$\mathcal{M}=1.188M_{\odot}$}
	\end{subfigure}
	~ %add desired spacing between images, e. g. ~, \quad, \qquad, \hfill etc. 
	\begin{subfigure}[b]{0.98\textwidth}
		\includegraphics[width=\textwidth]{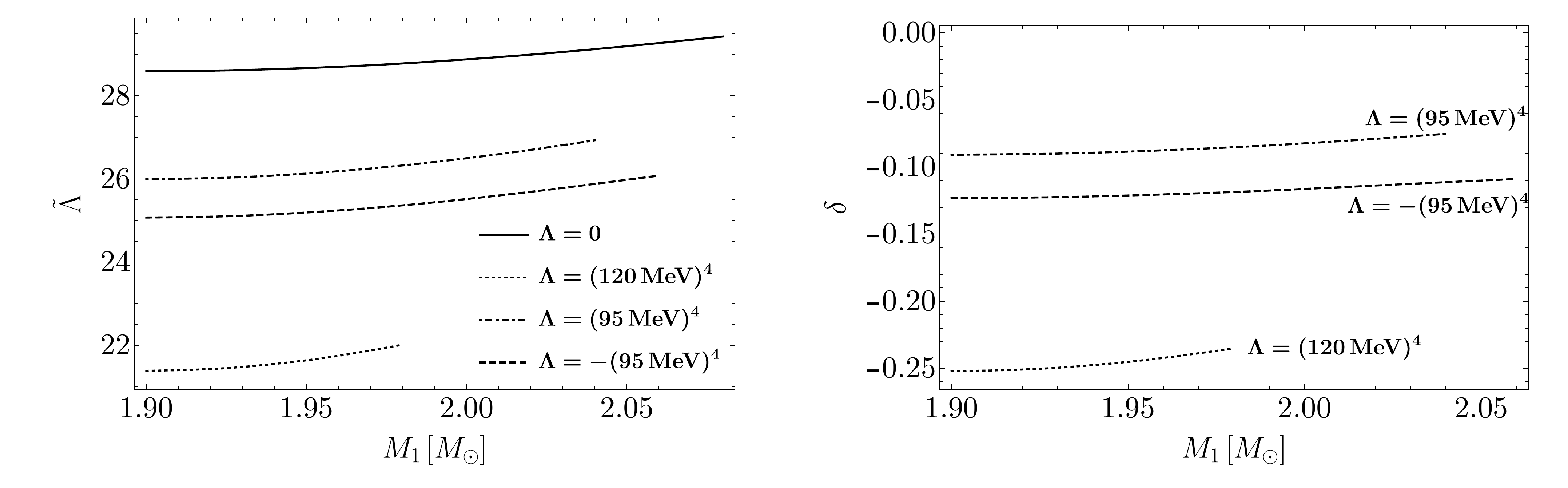}
		\caption{$\mathcal{M}=1.65M_{\odot}$}
	\end{subfigure}
	\caption{\small  Plot of the deviation of the combined dimensionless tidal deformability as a function of the heavy star mass for the AP4 EoS with different values for the chirp mass.  Plots on the left show the value of $\tilde{\Lambda}$, while plots on the right show the fractional deviation, $\delta$. The chirp mass $\mathcal{M}=1.188M_{\odot}$ is the same as the one of GW170817, while $\mathcal{M}=1.65M_{\odot}$ corresponds to a chirp mass where if the two NS masses are equal they have a mass of $1.9M_{\odot}$. For the smaller chirp mass the effect is rather small, however for a higher chirp mass the effect can be as large as $25\%$. Again the $\alpha$-parameter of \eqref{jump} is chosen to be $\alpha=3$.}
	\label{fig:ap4-dev}
\end{figure}

\begin{figure}[h!]
	\centering
	\begin{subfigure}[b]{0.49\textwidth}
		\includegraphics[width=\textwidth]{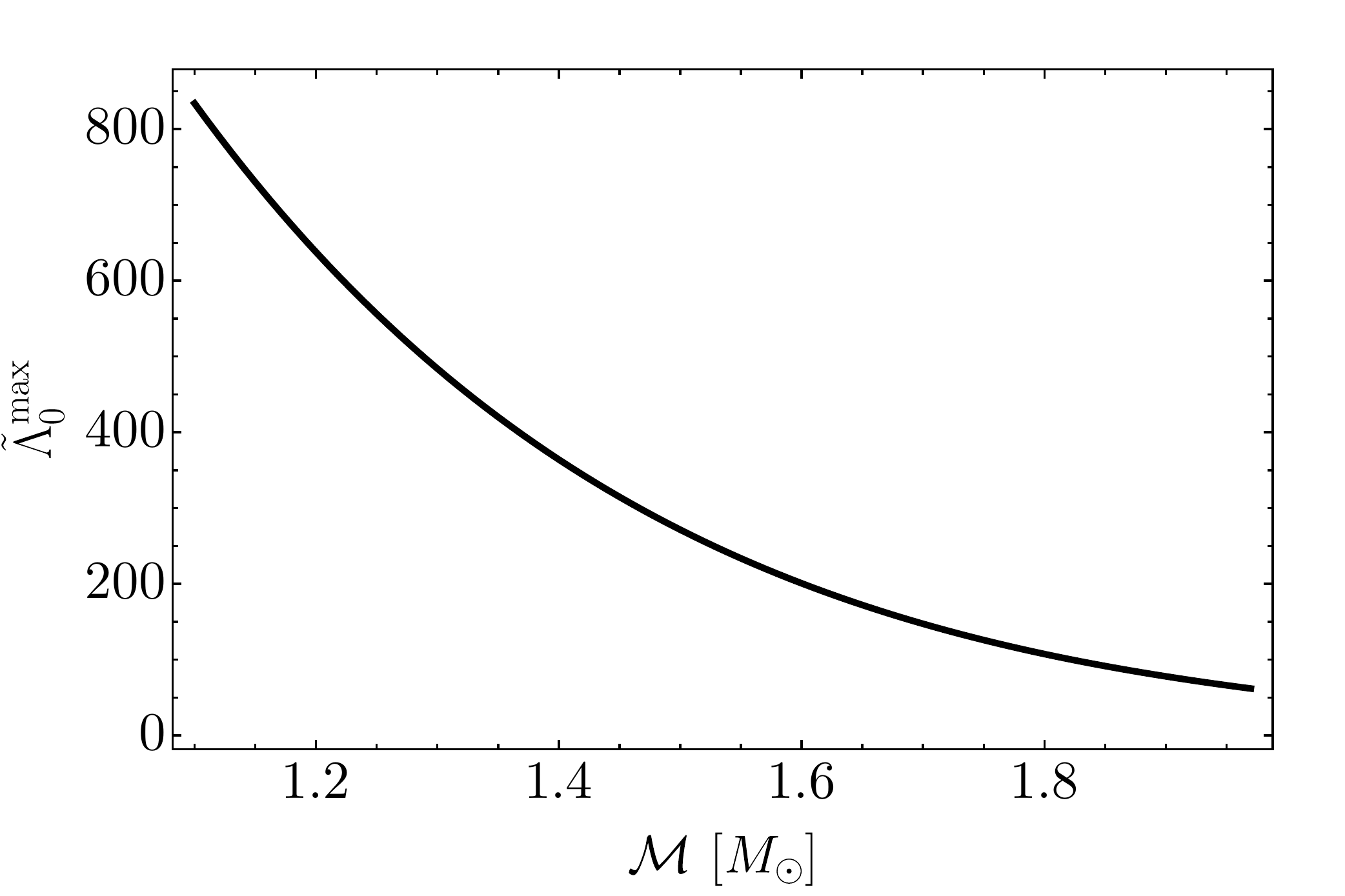}
	\end{subfigure}
	\begin{subfigure}[b]{0.49\textwidth}
		\includegraphics[width=\textwidth]{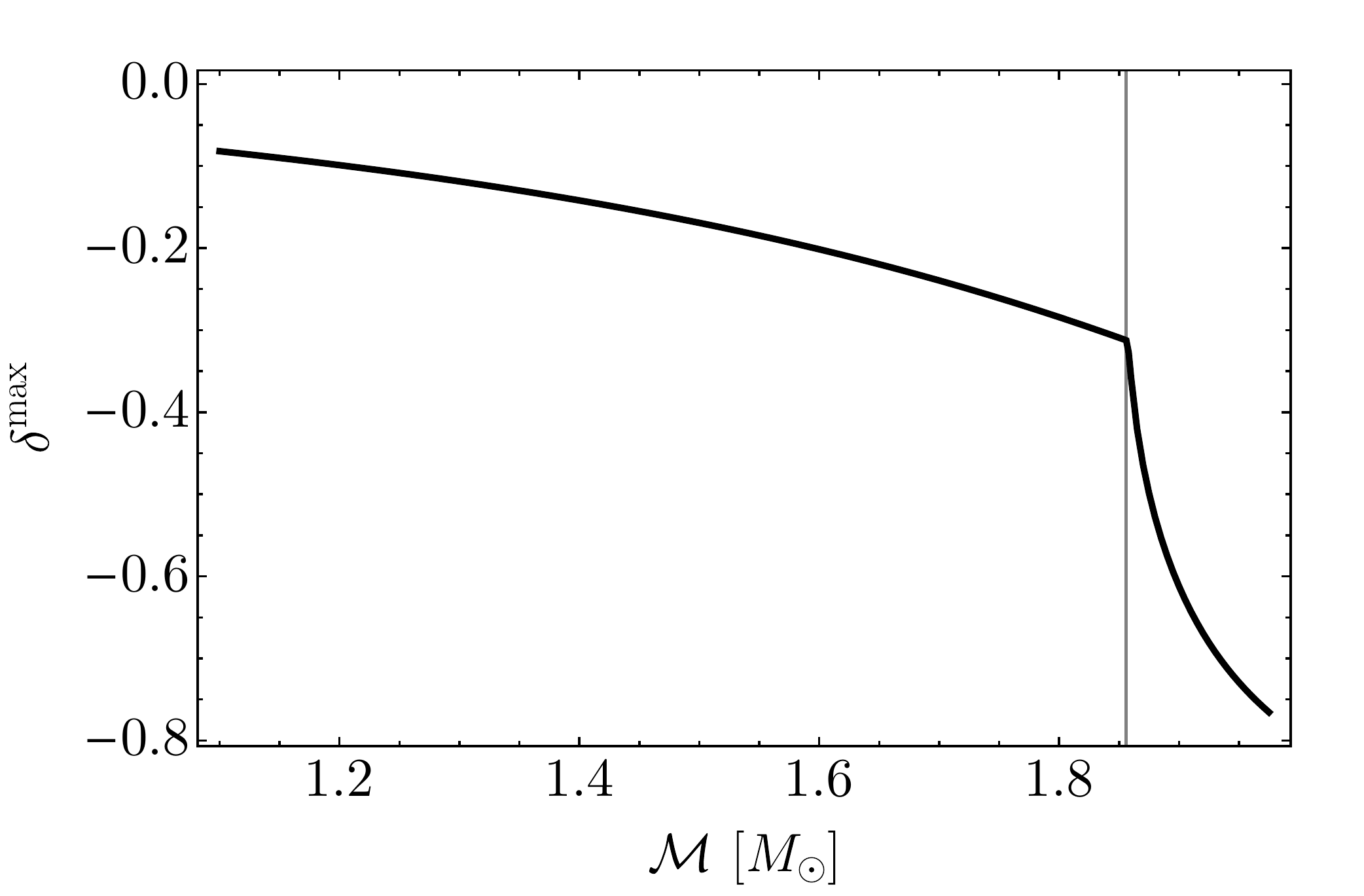}
	\end{subfigure}
	\caption{\small Dependence on the chirp mass in the Hebeler et al.\ parametrization, keeping the heaviest star mass fixed at $M_1=2.27M_\odot$ (the maximum value for the $\Lambda=(150\MeV)^4$ curve). The left plot shows the corresponding value of the combined tidal deformability for the $\Lambda=0$ curve. The right plot represents the relative deviation of the combined tidal deformability by turning on $\Lambda=(150\MeV)^4$ and is a measure of how the effect of VE potentially increases with the chirp mass.}
	\label{fig:maxvschirp}
\end{figure}

\begin{figure}[h!]
	\centering
	\begin{subfigure}[b]{0.48\textwidth}
		\includegraphics[width=\textwidth]{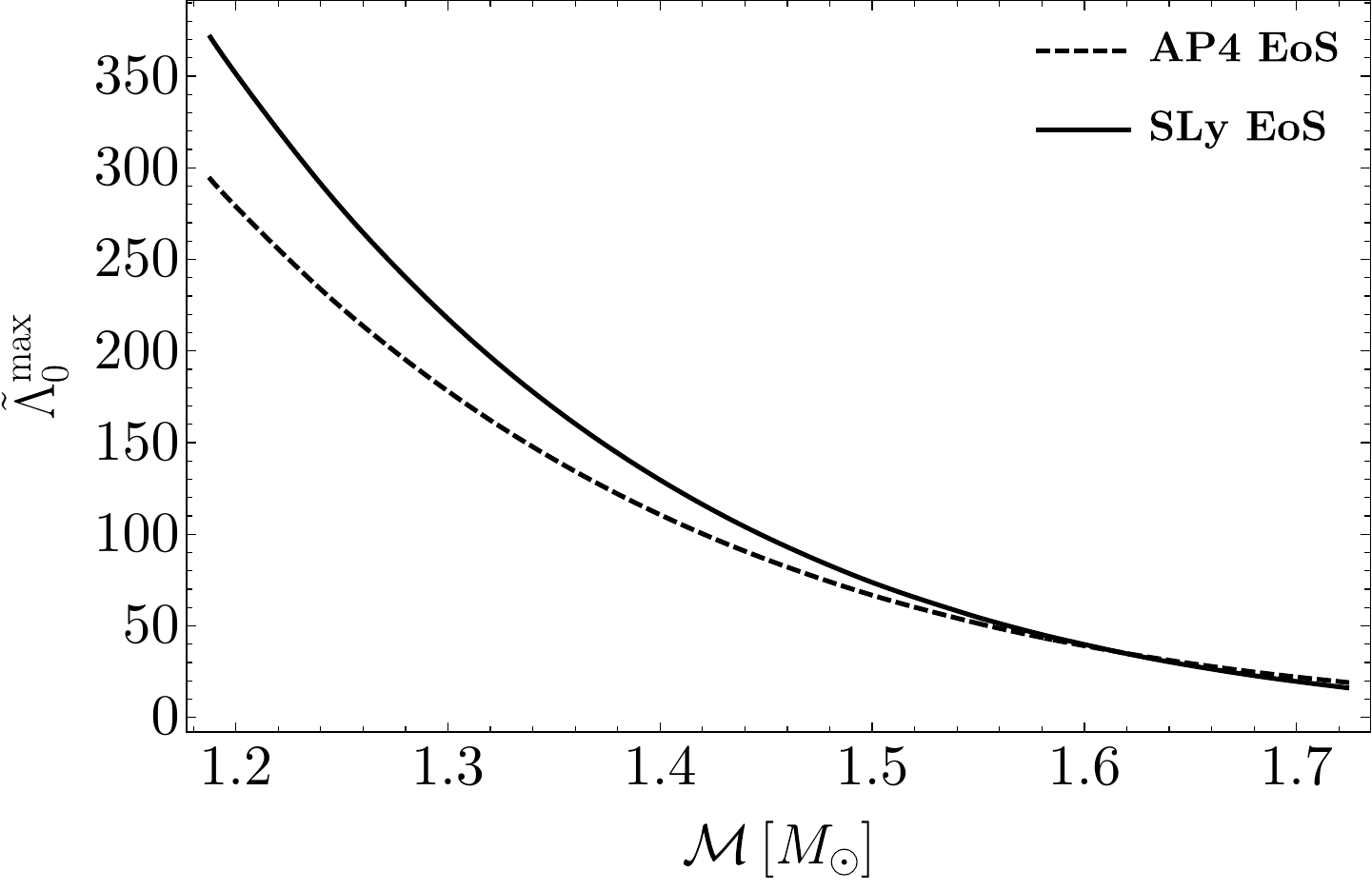}
	\end{subfigure}
	~ %add desired spacing between images, e. g. ~, \quad, \qquad, \hfill etc. 
	\begin{subfigure}[b]{0.48\textwidth}
		\includegraphics[width=\textwidth]{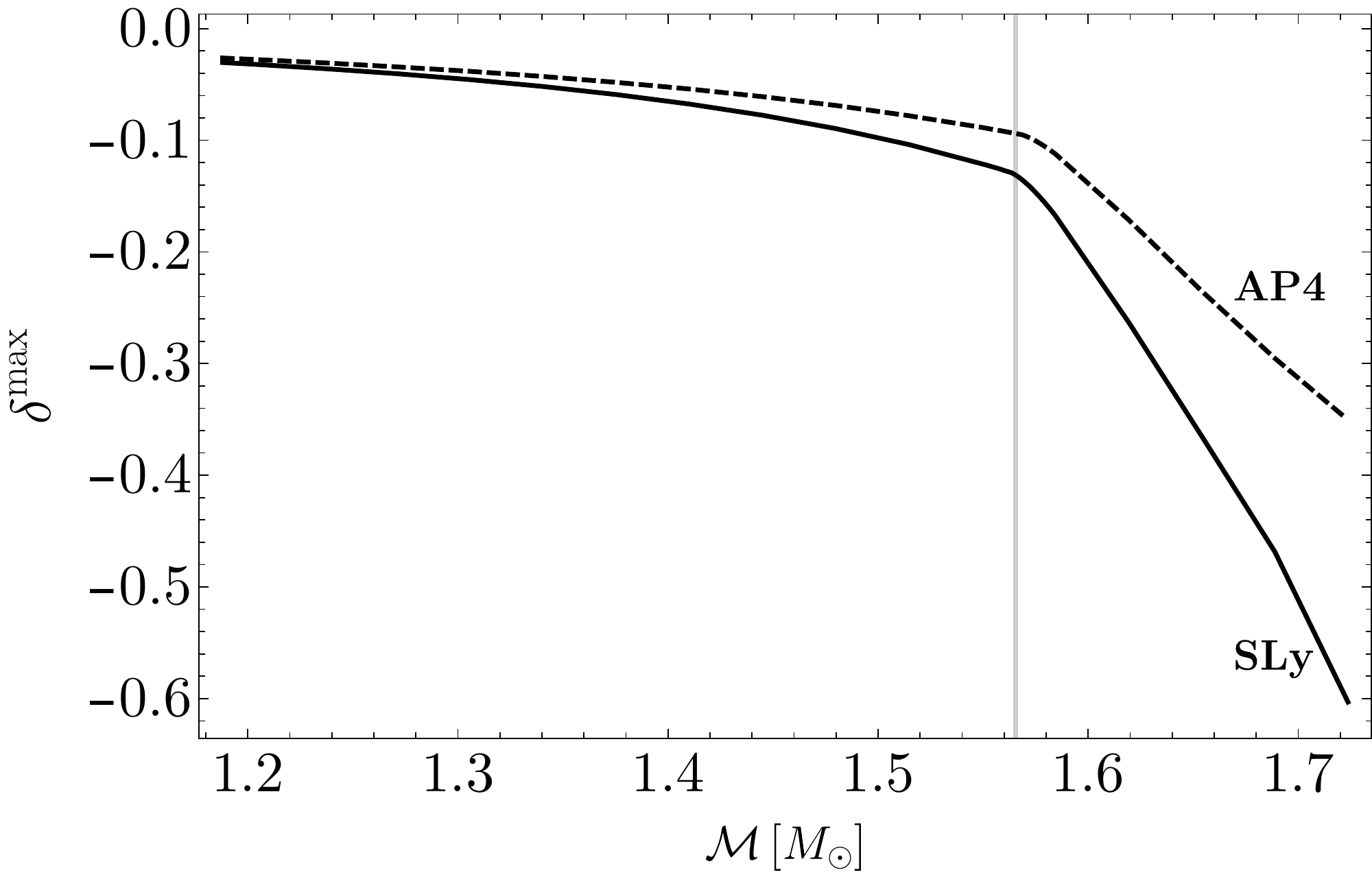}
	\end{subfigure}
	\caption{\small Dependence on the chirp mass in the AP4 and SLy parametrizations, keeping the heaviest star mass fixed at $M_1=1.98M_\odot$ (the maximum value for the $\Lambda=(120\MeV)^4$ curve).   The chirp mass range is from $\mathcal{M}=1.188M_{\odot}$ to $\mathcal{M}\approx 1.72M_{\odot}$, where the latter corresponds to the case when both stars have masses $M_{1,2}=1.98M_{\odot}$.  The left plot shows the corresponding value of the combined tidal deformability for the $\Lambda=0$ curves. The right plot represents the relative deviation of the combined tidal deformability and is a measure of how the effect of VE potentially increases with the chirp mass.  The vertical gray line denotes the chirp mass at which the light star mass reaches the critical mass for the phase transition.}
	\label{fig:lambda-tilde-vs-chirp}
\end{figure}

\begin{figure}[t]
	\centering
		\includegraphics[width=0.8\textwidth]{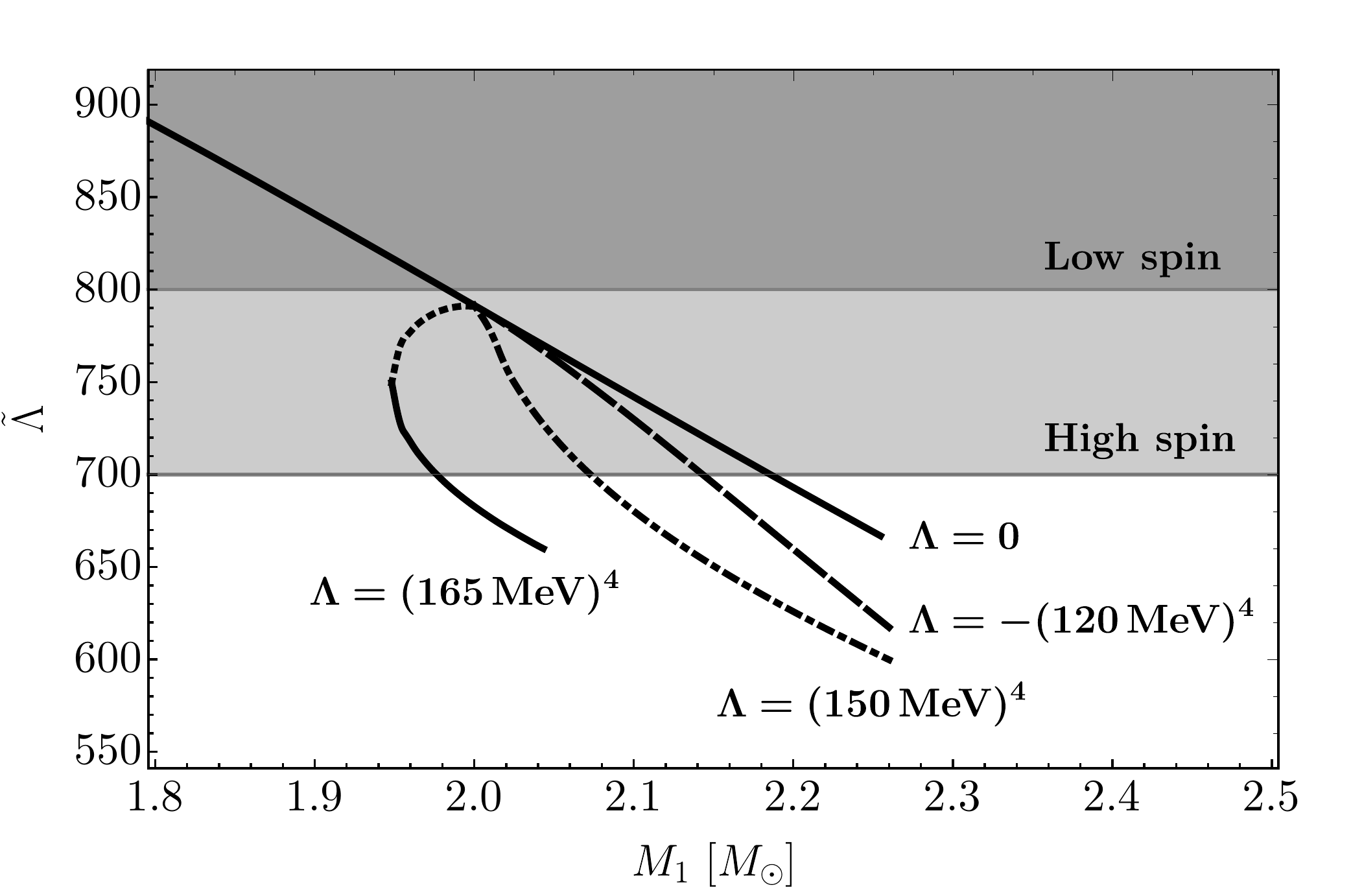}
	\caption{\small Combined tidal deformability $\tilde{\Lambda}$ as a function of the heavy star mass $M_1$ for the Hebeler et al.\ parametrization with $\alpha=3$. The chirp mass is the same as in the event GW170817. The figure shows the upper bounds set by the LIGO/Virgo analysis and demonstrates how a nonzero value of $\Lambda$ can affect the allowed mass range.}
	\label{fig:LIGOlimits}
\end{figure}

The rest of this section contains our results for the effects of VE on the tidal deformabilities, which will be presented in a series of plots. Each plot will be presented both for the Hebeler et al.\ EoS, which allows for larger NS masses and hence larger effects from VE, as well as for the AP4 and SLy EoS's, which cut NS masses off at $2 M_{\odot}$ and thus have smaller VE effects. 
\begin{itemize}
\item Fig.~\ref{fig:HebelerTidal} shows the individual tidal deformabilities of both NS's and the effect of VE using the Hebeler et al.\ EoS.

\item Fig.~\ref{fig:HebelerGWphase} translates the effects of VE into a fractional shift of the combined tidal deformability $\tilde{\Lambda}$. This plot shows that the effect of VE can be as large as 70\% and is generically sizable for the case of the Hebeler et al.\ EoS. 

\item Figs.~\ref{tidaldeform_ligo_slyap4}, \ref{fig:sly-dev}, \ref{fig:ap4-dev} repeat the same analyses for the AP4 and SLy EoS's, where we see that the deviations are generically smaller, but can still reach 25--30\% for larger chirp masses. 

\item Figs.~\ref{fig:maxvschirp}, \ref{fig:lambda-tilde-vs-chirp} emphasize the role of the chirp mass: they show the maximal achievable effect of VE as the chirp mass is increased. We can see that observing events with chirp masses above $1.6$--$1.8 M_\odot$ will be key to observing the effects of VE. 

\item Fig.~\ref{fig:LIGOlimits} shows the effect of VE on GW170817 (assuming the Hebeler EoS) where it can significantly change the allowed region of NS masses one would infer from the constraint on the tidal deformability.  

\end{itemize}

The EoS parametrization of Hebeler et al.~\cite{Hebeler:2013nza} allows for large possible deviations in the tidal deformability when a VE term is added to the central core in the new phase.  In Fig.~\ref{fig:HebelerTidal}, we show the effect of varying the VE term for a selection of 3 different input chirp masses.  The curves are obtained by fixing the chirp mass $\mathcal{M}$ at a few representative values and then scanning over the mass of the heaviest star, $M_1$.  Typically it is found that the heavier the star, the smaller the tidal deformability.  This is largely due to the fact that more massive stars typically have smaller radii, and thus respond less to external tidal fields.  

We note that the $\Lambda=(165\MeV)^4$ curve in the third plot is composed of two separate branches, corresponding to the two separate stable stars with equal masses but different radii as explained in the previous section.   The branch with the highest values of $\bar{\lambda}_2$ corresponds to only the most massive star laying in the disconnected branch of the $M(R)$ curve, while in the other case both stars in the binary would come from the disconnected branch. Part of the reason why the deviations from the $\Lambda=0$ curve are significant here can be found directly in Fig.~\ref{fig:HebelerMR}.  Since there the maximum mass for the $\Lambda=0$ curve is close to $3M_\odot$, the curves that correspond to a nonzero $\Lambda$ can depart significantly without being excluded by the measurement of the most massive neutron star, $(2.01\pm0.04)M_\odot$.  
%The heavy stars have a larger fraction of their total mass in VE, leading to a larger effect when varying its magnitude.

As we are most interested in the changes brought about by considering non-vanishing VE, it is useful to introduce a variable that quantifies the relative shift in $\tilde{\Lambda}$ due to VE:
\beq
\label{eq:gw_phase_dev}
\delta \equiv \frac{\tilde{\Lambda}-\tilde{\Lambda}_0}{\tilde{\Lambda}_0}~,
\eeq
where $\tilde{\Lambda}_0$ is the value of $\tilde{\Lambda}$ obtained by taking the VE term to zero. 

The deviation as a function of the heavy star mass, $M_1$, for the Hebeler et al.\ parametrization is shown in Fig.~\ref{fig:HebelerGWphase}. The negative values for $\delta$ mean that introducing a VE term lowers the value of $\tilde{\Lambda}$.   In order to isolate as much as possible the effects that a nonzero value of $\Lambda$ has on the internal structure of the stars, we are comparing each point in a given curve with the corresponding event on the $\Lambda=0$ curve that has the same neutron star masses. Therefore, any deviation in the value of $\tilde{\Lambda}$ comes entirely from the change in the tidal deformabilities, $\bar{\lambda}_i$.

Even with the more conservative SLy and AP4 models one still finds large deviations in $\tilde{\Lambda}$ for events with larger chirp masses.  The case of the (smaller) chirp mass corresponding to GW170817 is displayed in Fig.~\ref{tidaldeform_ligo_slyap4}, and the  deviations in deformability are small.  This is because the combined deformability is typically dominated by the contribution from the less massive star, which does not contain a core in the new phase where VE plays a role. However for higher chirp masses the effect of vacuum energy can be sizable even for the SLy and AP4 EoS's, as shown in Figs.~\ref{fig:sly-dev} and \ref{fig:ap4-dev}. As the chirp mass  increases more of the star contains the new phase, and eventually both stars typically contain cores in the new phase, yielding the increased sensitivity to VE.  The high chirp mass we have chosen for these figures corresponds, if the stars are of equal mass, to individual masses of $1.9 M_\odot$, approaching that of the most massive NS observed to date. One can see that for this case the deviation can be as large as $37\%$, even for these more conservative equations of state.

%The small changes in the deformabilities for GW170817 implies that 
%for these more conservative equations of state results from this event are not sensitive to the presence of VE. However, future improvements in sensitivity may eventually yield observable deviations from the Newtonian limit that give nonzero values for these parameters, particularly if a population of mergers with larger chirp mass appear in future data taking runs.
%The tidal deformabilities for SLy and AP4 equations of state with different choices of VE are shown in Fig.~\ref{tidaldeform_ligo_slyap4}. All curves are obtained by varying the mass of the heavier star with the chirp mass is fixed to be $\mathcal{M}=1.188 M_{\odot}$. For these models, the variation due to the nonzero VE is somewhat small. The reason for this is that the lighter of the two stars, which dominates the contribution to $\tilde{\Lambda}$, does not have a core in the novel phase of QCD.  Changing VE only has an effect through the more massive star, which has a smaller value of $\bar{\lambda}$.

%In Figures.~\ref{fig:maxvschirp} and \ref{fig:lambda-tilde-vs-chirp}, we show the behavior of $\tilde{\Lambda}$ as a function of chirp mass for the Hebeler et al.\ EoS and the AP4/SLy EoS respectively, for representative values of VE.  The mass of the heavier of the two stars is held fixed in these curves.  These plots clearly demonstrate that if deformabilities in mergers with larger chirp mass can be constrained in the future, significant limits on the effect of VE in high density QCD plasma could be placed.

Since the chirp mass is the most accurately measured property of the NS merger, it is worthwhile to 
examine the dependence of $\delta$ (characterizing the sensitivity to VE) on the chirp mass. In Figs.~\ref{fig:maxvschirp} (Hebeler) and \ref{fig:lambda-tilde-vs-chirp}  (AP4 and SLy) we plot $\tilde{\Lambda}_0^\textnormal{max}$ and $\delta^\textnormal{max}$ as a function of the chirp mass. The superscript expresses the fact that, when evaluating the quantities in Eqs.~\eqref{eq:lambdatilde} and \eqref{eq:gw_phase_dev}, the mass of the heavy star is kept fixed at the maximal value allowed by the corresponding fixed value of $\Lambda$. Fixing one of the stars to have maximal mass will generically (though not always) give the largest VE effect on $\tilde{\Lambda}$.  The important result of the plots is that the deviation increases substantially above a certain chirp mass denoted by the vertical gray line in the plots. This threshold corresponds to the chirp mass for which the lighter star mass also reaches the critical mass for the phase transition. Therefore, the large deviation can be understood from the fact that both stars are in the new phase. 

In our final plot in Fig.~\ref{fig:LIGOlimits}, we display the limits that GW170817 places on VE assuming the Hebeler et al.\ parametrization.  In particular, we note that including a VE term significantly changes the allowed range of the individual masses for the high-spin assumption.  The effect is less pronounced for the SLy and AP4 models.  As more data on NS mergers are collected with some of those corresponding to mergers of more massive stars,  strong limits could be placed on the EoS of dense nuclear matter. This will especially be true once the sensitivities for
probing the tidal contributions to the gravitational wave phase further improve. 

The future outlook is difficult to extrapolate, but promising.  The constraints that are placed in the coming years will depend strongly on currently uncertain characteristics of NS binaries or NS-black hole mergers that will be captured by upcoming data-taking runs at LIGO and other GW observatories~\cite{Kumar:2016zlj}.  Constraints will depend upon masses, spins, and branch populations in cases where there are multiple configurations with the same mass.  In addition, the sensitivities of the experiments will evolve, and may be able to better capture higher order contributions to the waveforms.  Finally, utilization of neutron stars as laboratories to study very high density physics and VE depends crucially on a precise theoretical calculation of the QCD equation of state at high densities \cite{Rezzolla:2016nxn}.  Taking an optimistic viewpoint on these issues leads us to the conclusion that neutron star mergers can tell us about the interface of gravity and quantum field theory in a regime never before tested.

%%%%%%%%%%%%%%%%%%%%%%%%%%%%%%%%%%%%%%%%%%%%%%%%
%%%%%%%%%%%%%%%%%%%%%%%%%%%%%%%%%%%%%%%%%%%%%%%%
 \section{Conclusions}
 
 In this paper, we have argued that neutron star mergers can be a valuable tool for testing new phases of QCD at large densities, in particular for finding the contribution of a VE term in exotic high density phases. To study the effects of such a new phase on neutron star observables, we have started with the conventional $7$-layer parametrization of the EoS, then assumed a nonzero value for the VE in the innermost layer leading to a jump in the energy density.
 For the three benchmark models we have chosen, we have calculated the $M(R)$ curves and tidal Love numbers for different values of the VE. By using those results, we have obtained individual tidal deformabilities and calculated the combined dimensionless tidal deformability parameter which can be constrained by neutron star mergers observed in gravitational wave observatories. We have found that for larger chirp masses, the nonzero VE at the innermost core can have an $\mathcal{O}(1)$ effect on the combined dimensionless tidal deformability parameter, hence future observations of neutron star merger chirps can place strong limits on the EoS of dense nuclear matter. We have also shown that for some parameters, introducing a nonzero VE can create a disconnected branch of stable neutron star solutions allowing the possibility of having two neutron stars of the same mass with significantly different radii. This possibility is unique to EoS's which have a  phase transition at the core, hence it is a smoking gun for new phases of QCD.

%%%%%%%%%%%%%%%%%%%%%%%%%%%%%%%%%%%%%%%%%%%%%%%%
%%%%%%%%%%%%%%%%%%%%%%%%%%%%%%%%%%%%%%%%%%%%%%%%
\section*{Acknowledgments}
\setcounter{equation}{0}
\setcounter{footnote}{0}

We thank Brando Bellazzini and Javi Serra for many helpful conversations, and for collaboration at early stages of this work.  We also thank Mark Alford for a useful discussion.  J.H.\ thanks the Aspen Center for Physics for its hospitality while some of this work was performed.   C.C., C.E., and G.R.\ thank the Galileo Galilei Institute of the INFN in Florence, Italy for its hospitality while this work was in progress. C.E, J.H., and G.R.\ thank Cornell University for hospitality throughout this work. C.C.\ is supported in part by the NSF grant PHY-1719877. C.E., J.H., and G.R.\ are supported in part by the DOE under grant DE-FG02-85ER40237. J.T.\ was supported in part by the DOE under grant DE-SC-000999.

%\appendix
%%%%%%%%%%%%%%%%%%%%%%%%%%%%%%%%%%%%%%%%%%%%%%%%
%%%%%%%%%%%%%%%%%%%%%%%%%%%%%%%%%%%%%%%%%%%%%%%%
%\section*{Appendix}

%%%%%%%%%%%%%%%%%%%%%%%%%%%%%%%%%%%%%%%%%%%%%%%%
%%%%%%%%%%%%%%%%%%%%%%%%%%%%%%%%%%%%%%%%%%%%%%%%

\end{document}